\documentclass[12pt,eqsecnum]{article}
\usepackage[dvips]{graphicx}
\usepackage{amssymb}
\usepackage{amsmath}
\usepackage{ulem}
\usepackage[dvipsnames]{xcolor}
\usepackage{soul}
\usepackage{ascmac}
\usepackage{cite}

\makeatletter

\@addtoreset{equation}{section}
\makeatletter

\newcommand{\qed}{\hbox{\rule[-2pt]{6pt}{6pt}}}
\newcommand{\D}{{\rm d}}

\newtheorem{Prop}{Proposition}

\newcommand{\dalm}{\kern1pt\vbox{\hrule height 0.9pt\hbox{\vrule width
0.9pt\hskip 2.5pt\vbox{\vskip 5.5pt}\hskip 3pt\vrule width 0.3pt}\hrule height
0.3pt}\kern1pt}

\begin{document}

\begin{titlepage}
\vfill
\begin{flushright}
\today
\end{flushright}

\vfill
\begin{center}
\baselineskip=16pt
{\Large\bf 
Energy conditions for non-timelike thin shells\\
}
\vskip 0.5cm
{\large {\sl }}
\vskip 10.mm
{\bf Hideki Maeda${}^{a,b}$} \\

\vskip 1cm
{
${}^a$ Department of Electronics and Information Engineering, Hokkai-Gakuen University, Sapporo 062-8605, Japan.\\
${}^b$ Max-Planck-Institut f\"ur Gravitationsphysik (Albert-Einstein-Institut), \\ Am M\"uhlenberg 1, D-14476 Potsdam, Germany.\\
\texttt{h-maeda@hgu.jp}
}
\vspace{6pt}
\end{center}
\vskip 0.2in
\par
\begin{center}
{\bf Abstract}
\end{center}
\begin{quote}
We study energy conditions for non-timelike thin shells in arbitrary $n(\ge 3)$ dimensions.
It is shown that the induced energy-momentum tensor $t_{\mu\nu}$ on a shell $\Sigma$ is of the Hawking-Ellis type I if $\Sigma$ is spacelike and either of type I, II, or III if $\Sigma$ is null.
Then, we derive simple equivalent representations of the standard energy conditions for $t_{\mu\nu}$.
In particular, on a spacelike shell or on a null shell with non-vanishing surface current, $t_{\mu\nu}$ inevitably violates the dominant energy condition.
If the surface pressure on the null shell is vanishing in addition, $t_{\mu\nu}$ is of type III and violates all the standard energy conditions.
Those fully general results are obtained without imposing a spacetime symmetry and can be used in any theory of gravity.
Lastly, several applications of the main results are presented in general relativity in four dimensions.
\vfill
\vskip 2.mm
\end{quote}
\end{titlepage}




\tableofcontents

\newpage

\section{Introduction}

In gravitation physics, a thin shell $\Sigma$ is a junction hypersurface of two spacetimes on which there is a non-vanishing induced energy-momentum tensor $t_{\mu\nu}$.
Possible embeddings of $\Sigma$ in a given set of bulk spacetimes and the resulting $t_{\mu\nu}$ are determined by the {\it junction conditions}.
The first junction conditions require that the induced metrics on both sides of $\Sigma$ are the same.
Then, the second junction conditions relate $t_{\mu\nu}$ and the jump of the extrinsic curvature (transverse curvature) of $\Sigma$ if $\Sigma$ is non-null (null).
In general relativity, the second junction conditions are derived from the Einstein equations and referred to as the Israel junction conditions for non-null $\Sigma$~\cite{Israel:1966rt} and the Barrab\`{e}s-Israel junction conditions for null $\Sigma$~\cite{Barrabes:1991ng}.
The Barrab\`{e}s-Israel junction conditions have been reformulated by Poisson to provide a simple characterization of the thin-shell energy-momentum tensor $t_{\mu\nu}$~\cite{Poisson:2002nv}.
(Sec.~3 in the textbook~\cite{Poisson} summarizes the junction conditions in general relativity.)

In fact, in order to grasp the essence of general relativistic gravitational phenomena, thin shells have been used to construct simple models in a very wide variety of contexts.
It is not possible to list all the papers, but examples of such phenomena are gravitational collapse~\cite{Hajicek:1992pu,Echeverria:1993wf,Crisostomo:2003xz,Cardoso:2016wcr}, growth of cosmic voids~\cite{Sakai:1993vn,Maeda:2011yq,Sakai:1999xx} and bubbles~\cite{Berezin:1982ur,Berezin:1987bc,Blau:1986cw,Sato:1981bf,Maeda:1981gw,Sato:1981gv,Kodama:1981gu}, and brane cosmology~\cite{Randall:1999ee,Randall:1999vf,Shiromizu:1999wj,Kraus:1999it,Ida:1999ui}.
An exact model for the mass-inflation instability of the inner horizon of a charged black hole has also been constructed using a null shell~\cite{Ori:1991zz}.
The junction conditions have also been established in a large class of scalar-tensor theories of gravity for non-null $\Sigma$~\cite{Sakai:1992ud,Barcelo:2000js,Padilla:2012ze} as well as for general $\Sigma$~\cite{Aviles:2019xae}.
In higher curvature theories, they have been formulated for non-null $\Sigma$ in Einstein-Gauss-Bonnet gravity~\cite{Davis:2002gn,Gravanis:2002wy} and Lovelock gravity~\cite{Miskovic:2007mg}. 
(See also Sec.~15 in the textbook~\cite{Padmanabhan:2010zzb}.)
However, the junction conditions for null $\Sigma$ in such theories are still missing.
(See~\cite{j-conditions} for recent developments in the research of junction conditions.)

For such a shell model to be physically reasonable, $t_{\mu\nu}$ induced on the shell should satisfy at least some of the standard energy conditions~\cite{Hawking:1973uf,Maeda:2018hqu}.
(See~\cite{Curiel:2014zba} for a nice review of the energy conditions.)
In arbitrary $n(\ge 3)$ dimensions, an energy-momentum tensor $T_{\mu\nu}$ is classified into the four Hawking-Ellis types~\cite{Maeda:2018hqu,Santos:1994cs,srt1995,hrst1996,rst2004} and equivalent representations of the standard energy conditions in terms of the orthonormal components of $T_{\mu\nu}$ in the canonical frame are available for each type~\cite{Martin-Moruno:2017exc,Maeda:2018hqu}.
Nevertheless, finding the canonical orthonormal frame by local Lorentz transformations is often not easy.
Accordingly, we derived equivalent representations of the standard energy conditions in the case where the orthonormal components of $T_{\mu\nu}$ admit only a single off-diagonal ``space-time'' component~\cite{Maeda:2022vld}.
Relations between the first junction conditions and the energy conditions were discussed in~\cite{Marolf:2005sr} for a timelike $\Sigma$.
If a shell $\Sigma$ is timelike, due to the Lorentzian signature on $\Sigma$, the energy conditions for $t_{\mu\nu}$ induced on the shell have to be examined in each case individually using those results.
In contrast, the situation is drastically simplified if a shell is non-timelike.
In this paper, we will present simple representations of the standard energy conditions for $t_{\mu\nu}$ on a non-timelike $\Sigma$ embedding in an $n(\ge 3)$-dimensional spacetime.

The present article is organized as follows. 
In the next section, after reviewing the standard energy conditions, the Hawking-Ellis classification of an energy-momentum tensor, and the junction conditions, we will present our main results.
In Sec.~\ref{sec:apply}, we will apply the main results in several physical situations in general relativity in four dimensions.
Throughout this article, the signature of the Minkowski spacetime is $(-,+,\ldots,+)$, and Greek indices run over all spacetime indices.
Other types of indices will be specified in the main text.
We adopt the units such that $c=1$ and use $\kappa_n:=8\pi G_n$ instead of the $n$-dimensional gravitational constant $G_n$.
The conventions of curvature tensors such as $[\nabla _\rho ,\nabla_\sigma]V^\mu ={R^\mu }_{\nu\rho\sigma}V^\nu$ and ${R}_{\mu \nu }={R^\rho }_{\mu \rho \nu }$.

\section{Energy conditions for thin-shells}
\label{sec:review}

We follow the definitions and notations adopted in~\cite{Maeda:2018hqu} for the energy conditions and the Hawking-Ellis types. 
In $n(\ge 3)$ dimensions, the standard energy conditions for an energy-momentum tensor $T_{\mu\nu}$ are stated as follows:
\begin{itemize}
\item {\it Null} energy condition (NEC): $T_{\mu\nu} k^\mu k^\nu\ge 0$ for any null vector $k^\mu$.
\item {\it Weak} energy condition (WEC): $T_{\mu\nu} v^\mu v^\nu\ge 0$ for any timelike vector $v^\mu$.
\item {\it Dominant} energy condition (DEC): $T_{\mu\nu} v^\mu v^\nu\ge 0$ and $J_\mu J^\mu\le 0$ hold for any timelike vector $v^\mu$, where $J^\mu:=-T^\mu_{\phantom{\mu}\nu}v^\nu$ is an energy-flux vector for an observer with its tangent vector $v^\mu$. 
\item {\it Strong} energy condition (SEC): $\left(T_{\mu\nu}-\frac{1}{n-2}Tg_{\mu\nu}\right) v^\mu v^\nu\ge 0$ for any timelike vector $v^\mu$.
\end{itemize}
The SEC is equivalent to the timelike convergence condition $R_{\mu\nu}v^{\mu}v^{\nu} \ge 0$ for any timelike vector in general relativity without a cosmological constant.

\subsection{Hawking-Ellis types}

An orthonormal frame is defined by a set of $n$ orthonormal basis vectors $\{{E}^\mu_{(\alpha)}\}$ $(\alpha=0,1,\cdots,n-1)$ that satisfy
\begin{equation}
{E}^\mu_{(\alpha)}{E}_{(\beta)\mu}=\eta_{(\alpha)(\beta)}=\mbox{diag}(-1,1,\cdots,1),
\end{equation}
which is equivalent to $g_{\mu\nu}=\eta_{(\alpha)(\beta)}E^{(\alpha)}_{\mu}E^{(\beta)}_{\nu}$.
The Minkowski metric $\eta_{(\alpha)(\beta)}$ in the orthonormal frame and its inverse $\eta^{(\alpha)(\beta)}$ are respectively used to lower and raise the indices $(\alpha)$.
Components of $T_{\mu\nu}$ in the orthonormal frame are given by $T_{(\alpha)(\beta)}=T_{\mu\nu}E_{(\alpha)}^{\mu}E_{(\beta)}^{\nu}$.
An orthonormal frame has a degree of freedom provided by local Lorentz transformations ${E}^\mu_{(\alpha)}\to{\tilde E}^\mu_{(\alpha)}:=L_{(\alpha)}^{~~(\beta)}{E}^\mu_{(\beta)}$, where $L_{(\alpha)}^{~~(\beta)}$ satisfies $L_{(\alpha)}^{~~(\gamma)}L_{(\beta)}^{~~(\delta)}\eta_{(\gamma)(\delta)}=\eta_{(\alpha)(\beta)}$.
$T_{(\alpha)(\beta)}$ behaves as a scalar under a diffeomorphism and as a two-tensor under a local Lorentz transformation.
We refer to such a mathematical object as a Lorentz-covariant tensor.
According to this terminology, a basis vector ${E}^\mu_{(\alpha)}$ is a Lorentz-covariant vector.

The Hawking-Ellis classification of $T_{\mu\nu}$ is performed according to the properties of the Lorentz-invariant eigenvalues $\lambda$ and eigenvectors $n^\mu$ (or Lorentz-covariant eigenvectors $n^{(\alpha)}=E^{(\alpha)}_\mu n^\mu$) that are determined by the following eigenvalue equations~\cite{Hawking:1973uf,Maeda:2018hqu}:
\begin{align}
T^{(\alpha)(\beta)} n_{(\beta)}=\lambda \eta^{(\alpha)(\beta)} n_{(\beta)}~~\Leftrightarrow~~T^{\mu\nu}n_\nu=\lambda g^{\mu\nu} n_\nu.\label{eigen-eq}
\end{align}
The characteristic equation to determine $\lambda$ is
\begin{align}
\det \left(T^{(\alpha)(\beta)}-\lambda \eta^{(\alpha)(\beta)}\right)=0.\label{chara-eq}
\end{align}
Since $n_{(\alpha)}n^{(\alpha)}=n_\mu n^\mu$ holds, a Lorentz-covariant vector $n^{(\alpha)}$ is referred to as timelike, spacelike, and null if $n_{(\alpha)}n^{(\alpha)}$ is negative, positive, and zero, respectively.

In three or higher dimensions ($n\ge 3$), $T_{\mu\nu}$ is classified into four types as summarized in Table~\ref{table:scalar+1}~\cite{Hawking:1973uf,Maeda:2018hqu}.
By local Lorentz transformations, we can write each type of $T^{(\alpha)(\beta)}$ in a canonical form, and then equivalent representations of the standard energy conditions are available~\cite{Maeda:2018hqu}.
\begin{table}[htb]
\begin{center}
\caption{\label{table:scalar+1} Eigenvectors of type-I--IV energy-momentum tensors.}
\scalebox{1.0}{
\begin{tabular}{|c|c|c|c|}
\hline
Type & Eigenvectors \\ \hline\hline
I & 1 timelike, $n-1$ spacelike \\ \hline
II & 1 null (doubly degenerated), $n-2$ spacelike \\ \hline
III & 1 null (triply degenerated), $n-3$ spacelike \\ \hline
IV & 2 complex, $n-2$ spacelike \\ 
\hline
\end{tabular} 
}\end{center}
\end{table}

\noindent
{\bf Type I}

The canonical form of type I is
\begin{equation}
\label{T-typeI}
T^{(\alpha)(\beta)}=\mbox{diag}(\rho,p_1,p_2,\cdots,p_{n-1})
\end{equation}
for which the characteristic equation (\ref{chara-eq}) gives 
\begin{equation} 
(\lambda+\rho)(\lambda-p_1)\cdots(\lambda-p_{n-1})=0,
\end{equation}
so that the eigenvalues are $\lambda=\{-\rho,p_1,p_2,\cdots,p_{n-1}\}$.
The eigenvector of $\lambda=-\rho$ is timelike and other eigenvectors are spacelike.
The standard energy conditions are equivalent to the following inequalities:
\begin{align}
\mbox{NEC}:&~~\rho+p_i\ge 0~~\mbox{for}~~i=1,2,\cdots,n-1,\label{NEC-I}\\
\mbox{WEC}:&~~\rho\ge 0\mbox{~in addition to NEC},\label{WEC-I}\\
\mbox{DEC}:&~~\rho-p_i\ge 0~~\mbox{for}~~i=1,2,\cdots,n-1\mbox{~in addition to WEC},\label{DEC-I}\\
\mbox{SEC}:&~~(n-3)\rho+\mbox{$\sum_{j=1}^{n-1}$}p_j\ge 0~~\mbox{~in addition to NEC.}\label{SEC-I}
\end{align}

\noindent
{\bf Type II}

The canonical form of type II is
\begin{equation} 
\label{T-typeII}
T^{(\alpha)(\beta)}=\left( 
\vphantom{\begin{array}{c}1\\1\\1\\1\\1\\1\end{array}}
\begin{array}{cccccc}
\rho+\nu &\nu&0&0&\cdots &0\\
\nu&-\rho+\nu&0&0&\cdots &0\\
0&0&p_2&0&\cdots&0 \\
0&0&0&\ddots&\vdots&\vdots \\
\vdots&\vdots&\vdots&\cdots&\ddots&0\\
0&0&0 &\cdots&0&p_{n-1}
\end{array}
\right)
\end{equation}
with $\nu\ne 0$, for which the characteristic equation (\ref{chara-eq}) gives
\begin{equation} 
(\lambda+\rho)^2(\lambda-p_2)\cdots(\lambda-p_{n-1})=0,
\end{equation}
so that the eigenvalues are $\lambda=\{-\rho,p_2,\cdots,p_{n-1}\}$.
The Lorentz-covariant eigenvector $n_{(\alpha)}={\bar k}_{(\alpha)}$ of the doubly degenerate eigenvalue $\lambda=-\rho$ is null, while the Lorentz-covariant eigenvectors $n_{(\alpha)}=w_{i(\alpha)}$ ($i=2,3,\cdots,n-1$) of the eigenvalues $\lambda=p_i$ are spacelike.
${\bar k}_{(\alpha)}$ and $w_{i(\alpha)}$ are given by
\begin{equation} 
\label{eigen-vec-null}
{\bar k}_{(\alpha)}=\left( 
\vphantom{\begin{array}{c}1\\1\\1\\1\\1\end{array}}
\begin{array}{c}
-1\\
1\\
0\\
0\\
\vdots \\
0
\end{array}
\right),\quad
w_{2(\alpha)}=\left( 
\vphantom{\begin{array}{c}1\\1\\1\\1\\1\end{array}}
\begin{array}{c}
0\\
0\\
1\\
0\\
\vdots \\
0
\end{array}
\right),\quad \cdots,\quad
w_{n-1(\alpha)}=\left( 
\vphantom{\begin{array}{c}1\\1\\1\\1\\1\end{array}}
\begin{array}{c}
0\\
0\\
0\\
0\\
\vdots \\
1
\end{array}
\right),
\end{equation}
with which $T^{(\alpha)(\beta)}$ can be written as
\begin{align} 
T^{(\alpha)(\beta)}=\nu {\bar k}^{(\alpha)} {\bar k}^{(\beta)}-\rho\,\eta_2^{(\alpha)(\beta)}+\sum_{i=2}^{n-1}p_iw_{i}^{(\alpha)} w_i^{(\beta)},\label{typeII-decomp}
\end{align}
where $\eta_2^{(\alpha)(\beta)}:=\mbox{diag}(-1,1,0,\cdots,0)$.
The standard energy conditions are equivalent to the following inequalities:
\begin{align}
\mbox{NEC}:&~~\nu\ge 0\mbox{~and~}\rho+p_i\ge 0~~\mbox{for}~~i=2,3,\cdots,n-1,\label{NEC-II}\\
\mbox{WEC}:&~~\rho\ge 0\mbox{~in addition to NEC},\label{WEC-II}\\
\mbox{DEC}:&~~\rho-p_i\ge 0~~\mbox{for}~~i=2,3,\cdots,n-1\mbox{~in addition to WEC},\label{DEC-II}\\
\mbox{SEC}:&~~(n-4)\rho+\mbox{$\sum_{j=2}^{n-1}$}p_j\ge 0~~\mbox{~in addition to NEC.}\label{SEC-II}
\end{align}

\noindent
{\bf Type III}

The canonical form of type III is
\begin{equation} 
\label{T-typeIII}
T^{(\alpha)(\beta)}=\left( 
\vphantom{\begin{array}{c}1\\1\\1\\1\\1\\1\\1\end{array}}
\begin{array}{ccccccc}
\rho+\nu &\nu&\zeta&0&0&\cdots &0\\
\nu &-\rho+\nu&\zeta&0&0&\cdots &0\\
\zeta&\zeta&-\rho&0&0&\cdots &0\\
0&0&0&p_3&0&\cdots&0 \\
0&0&0&0&\ddots&\vdots&\vdots \\
\vdots&\vdots&\vdots&\vdots&\cdots&\ddots&0\\
0&0&0&0 &\cdots&0&p_{n-1}
\end{array}
\right)
\end{equation}
with $\zeta\ne 0$, for which the characteristic equation (\ref{chara-eq}) gives
\begin{equation} 
(\lambda+\rho)^3(\lambda-p_3)\cdots(\lambda-p_{n-1})=0,
\end{equation}
so that the eigenvalues are $\lambda=\{-\rho,p_3,\cdots,p_{n-1}\}$.
The eigenvector of the triply degenerate eigenvalue $\lambda=-\rho$ is null and other eigenvectors are spacelike.
Any type-III energy-momentum tensor violates all the standard energy conditions.

We note that $\nu$ in Eq.~(\ref{T-typeIII}) can be set to zero by local Lorentz transformations if and only if $\zeta$ is non-zero~\cite{Martin-Moruno:2017exc}.
Nevertheless, the expression~\eqref{T-typeIII} with non-vanishing $\nu$ admits a limit $\zeta\to 0$ to type II and may be useful to identify a type-III energy-momentum tensor in a given spacetime.

\noindent
{\bf Type IV}

The canonical form of type IV is
\begin{equation} 
\label{T-typeIV}
T^{(\alpha)(\beta)}=\left( 
\vphantom{\begin{array}{c}1\\1\\1\\1\\1\\1\end{array}}
\begin{array}{cccccc}
\rho &\nu&0&0&\cdots &0\\
\nu&-\rho&0&0&\cdots &0\\
0&0&p_2&0&\cdots&0 \\
0&0&0&\ddots&\vdots&\vdots \\
\vdots&\vdots&\vdots&\cdots&\ddots&0\\
0&0&0 &\cdots&0&p_{n-1}
\end{array}
\right)
\end{equation}
with $\nu\ne 0$, for which the characteristic equation (\ref{chara-eq}) gives
\begin{equation} 
[(\lambda+\rho)^2+\nu^2](\lambda-p_2)\cdots(\lambda-p_{n-1})=0,
\end{equation}
so that the eigenvalues are $\lambda=\{-\rho+ i\nu,-\rho - i\nu,p_2,\cdots,p_{n-1}\}$.
The eigenvectors of the complex eigenvalues $\lambda=-\rho\pm i\nu$ are complex and other eigenvectors are spacelike.
Any type-IV energy-momentum tensor violates all the standard energy conditions.

We note that a canonical form of $T^{(\alpha)(\beta)}$ in the textbook~\cite{Hawking:1973uf} is different from Eq.~(\ref{T-typeIV}).
However, the expression (\ref{T-typeIV}) may be more useful as pointed out in~\cite{Martin-Moruno:2017exc}.

\subsection{Energy-momentum tensor of thin-shells}

For junction conditions, we follow the definitions and notations adopted in~\cite{Aviles:2019xae}.
We consider an $(n-1)$-dimensional junction hypersurface $\Sigma$ between two $n(\ge 3)$-dimensional spacetime regions $({\cal M}_+, g_{\mu\nu}^+)$ and $({\cal M}_-, g_{\mu\nu}^-)$.
The metric $g_{\mu\nu}^\pm$ is expressed in the coordinates $x_\pm^\mu$ on $({\cal M}_\pm, g_{\mu\nu}^\pm)$.
In the following subsections, we identify an $n$-dimensional spacetime $({\cal M},g_{\mu\nu})$ with the line element
\begin{align}
\D s_n^2=&g_{\mu\nu}(x)\D x^\mu \D x^\nu \label{bulk}
\end{align}
as $({\cal M}_+, g_{\mu\nu}^+)$ or $({\cal M}_-, g_{\mu\nu}^-)$ and its boundary as $\Sigma$.
Suppose that $\Sigma$ is described by $\Phi(x)=$constant in the bulk spacetime (\ref{bulk}).
We set the same intrinsic coordinates $y^a$ on both sides of $\Sigma$ and introduce the standard notation $[X]$ defined by 
\begin{equation}
[X]:= X^+-X^-,
\end{equation}
where $X^\pm$ are $X$'s evaluated either on the $+$ or $-$ side of $\Sigma$.

\subsubsection{Non-null shell}
\label{sec:non-null}

\begin{figure}[htbp]
\begin{center}
\includegraphics[width=0.7\linewidth]{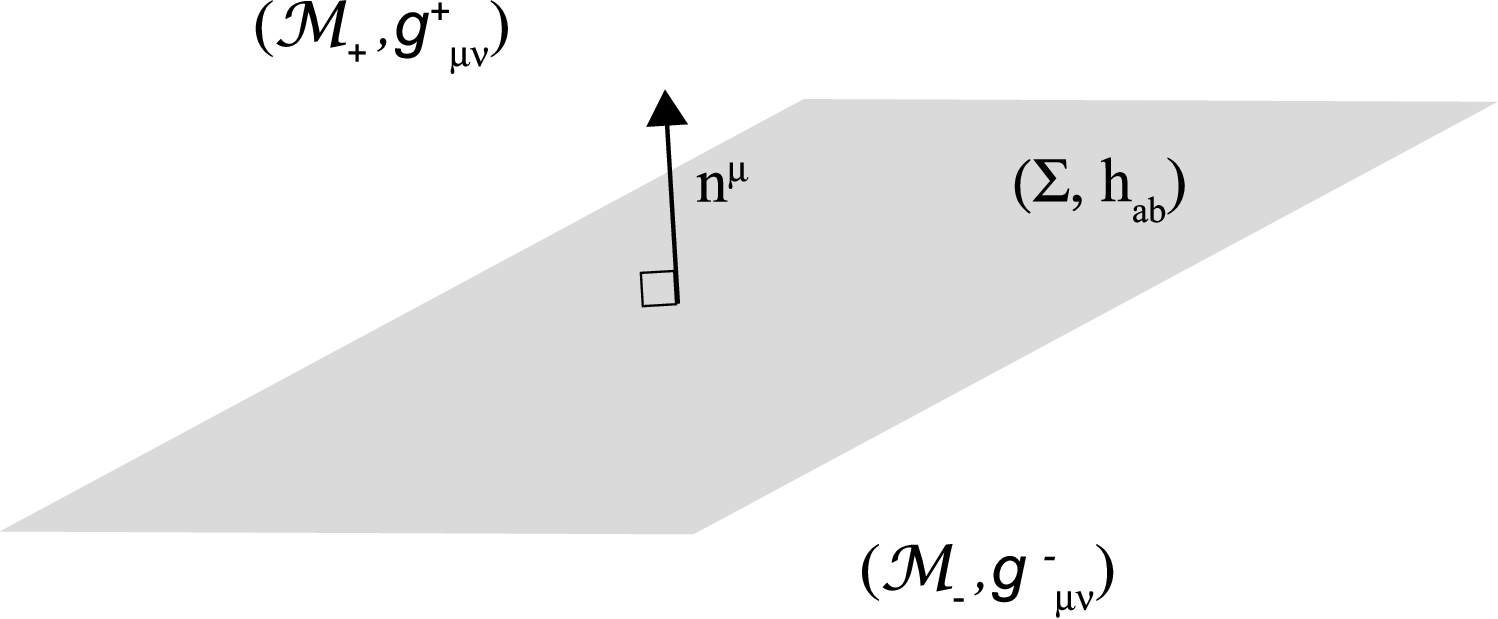}
\caption{\label{Fig-Nonnull-hypersurface} A non-null hypersurface $\Sigma$ partitions a spacetime into two regions ${\cal M}_+$ and ${\cal M}_-$.}
\end{center}
\end{figure}

Here we consider the case where $\Sigma$ is non-null as shown in Fig.~\ref{Fig-Nonnull-hypersurface}.
A unit normal vector $n^\mu$ to $\Sigma$ is given by 
\begin{align} 
n_\mu:=\frac{\varepsilon \nabla_\mu \Phi}{(\varepsilon g^{\rho\sigma}\nabla_\rho \Phi\nabla_\sigma \Phi)^{1/2}} \label{n-def}
\end{align} 
and satisfies $n^\mu n_\mu=\varepsilon$, where $\varepsilon=1$ ($-1$) corresponds to the case where $\Sigma$ is a timelike (spacelike) hypersurface. 
We choose $n^\mu$ to point from ${\cal M}_-$ to ${\cal M}_+$.

In the bulk spacetime $({\cal M},g_{\mu\nu})$, $\Sigma$ is described by $x^\mu=x^\mu(y)$ and the line element on $\Sigma$ is given by 
\begin{align}
\D s_{\Sigma}^2=&h_{ab}(y)\D y^a \D y^b,
\end{align}
where the induced metric $h_{ab}$ on $\Sigma$ is defined by 
\begin{align}
h_{ab}(y):=&g_{\mu\nu} e^\mu_a e^\nu_b, \qquad e^\mu_a:=\frac{\partial x^\mu}{\partial y^a}.\label{def-hab-nonnull}
\end{align}
$h_{ab}$ and its inverse $h^{ab}$ are used to raise or lower Latin indices, respectively.
A projection tensor is defined by $h_{\mu\nu}:=g_{\mu\nu}-\varepsilon n_\mu n_\nu$ which satisfies $h_{\mu\nu}n^\nu=0$ and $h_{ab}=h_{\mu\nu}e^\mu_a e^\nu_b$ (and therefore $h_{\mu\nu}=h_{ab}e^a_\mu e^b_\nu$).
The extrinsic curvature (or the {second fundamental form}) $K_{\mu\nu}$ of $\Sigma$ and its trace are defined by 
\begin{align}
K_{\mu\nu}:=&h^{~\rho}_\mu h^{~\sigma}_\nu \nabla_{\rho}n_{\sigma}\left(\equiv\frac12{\cal L}_nh_{\mu\nu}\right), \label{def-exK} \\
K:=&g^{\mu\nu}K_{\mu\nu}=\nabla_\mu n^\mu,
\end{align}
where ${\cal L}_n$ is the Lie derivative with respect to $n^\mu$.
$K_{\mu\nu}$ is symmetric and tangent to $\Sigma$, so that $K_{\mu\nu}n^\nu=0$ holds.

The first junction conditions at $\Sigma$ are given by 
\begin{align}
[h_{ab}]=0, \label{first-non-null}
\end{align}
which means that the induced metric on $\Sigma$ is the same on both sides of $\Sigma$.
Under the first junction conditions, one obtains the second junction conditions from the gravitational field equations\footnote{Independent second junction conditions may also be obtained from the field equations for matter fields in the bulk spacetime ${\cal M}$. (See~\cite{Aviles:2019xae} in the case of a scalar field.)} that determine the induced energy-momentum tensor $t_{\mu \nu}$ on $\Sigma$.
$t_{\mu \nu}$ is symmetric and tangent to $\Sigma$, so that $t_{\mu \nu}n^\nu=0$ holds.
In general relativity, the second junction conditions are obtained from the Einstein equations $G_{\mu\nu}+\Lambda g_{\mu\nu}=\kappa_nT_{\mu\nu}$ and referred to as the {\it Israel junction conditions}~\cite{Israel:1966rt}, which are given by 
\begin{align}
&-\varepsilon\left([K_{\mu\nu}]-h_{\mu\nu}[K]\right)=\kappa_nt_{\mu \nu}~~\Leftrightarrow~~-\varepsilon \left([K_{ab}]-h_{ab}[K]\right)=\kappa_n t_{ab}, \label{j-eq-summary1}
\end{align}
where $K_{ab}:=K_{\mu\nu}e^\mu_a e^\nu_b$, $K\equiv K_{ab}h^{ab}$, and $t_{ab}:=t_{\mu\nu}e^\mu_a e^\nu_b$.
We wil use the following expression 
\begin{align}
K_{ab}=&(\nabla_\mu n_{\nu})e^\mu_a e^\nu_b \nonumber \\
=&-n_{\mu}e_{a,b}^\mu-\Gamma^\kappa_{\mu\nu}n_\kappa e_a^\mu e_b^\nu \, ,\label{K-def}
\end{align}
where we have used $n_{\mu}e_{a}^\mu=0$ at the last equality.

\subsubsection{Null shell}

\begin{figure}[htbp]
\begin{center}
\includegraphics[width=0.8\linewidth]{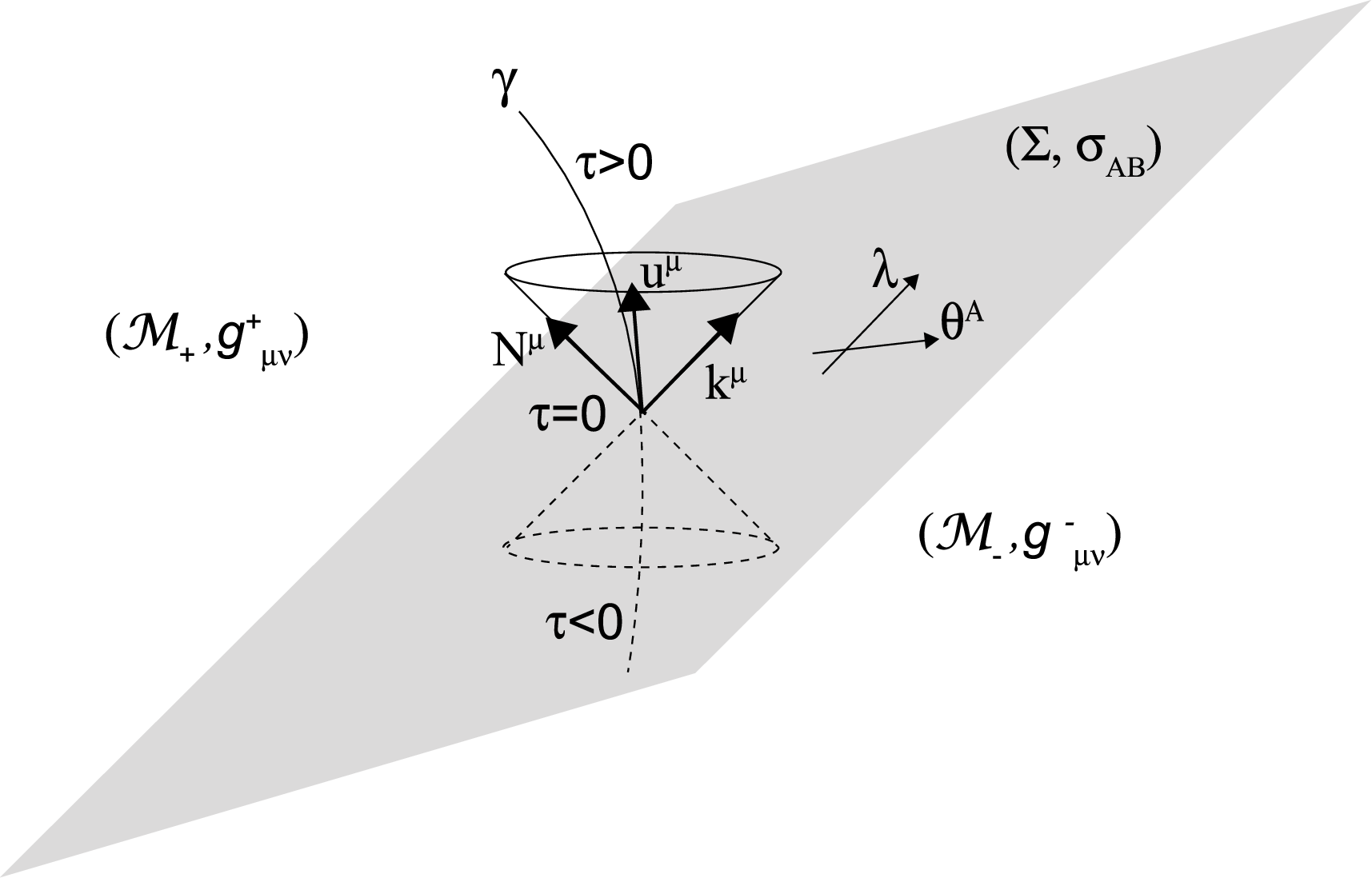}
\caption{\label{Fig-Null-hypersurface} A null hypersurface $\Sigma$ partitions a spacetime into two regions ${\cal M}_+$ and ${\cal M}_-$.}
\end{center}
\end{figure}

In the case where $\Sigma$ is null, our convention is such that ${\cal M}_-$ is in the past of $\Sigma$ and ${\cal M}_+$ is in the future as shown in Fig.~\ref{Fig-Null-hypersurface}.
Since the unit normal vector (\ref{n-def}) cannot be used for null $\Sigma$, we introduce a null vector $k^\mu$ defined by 
\begin{eqnarray}
k^\mu=-\nabla^\mu\Phi,\label{k-def-null}
\end{eqnarray}
which is tangent to the generators of $\Sigma$. (See Sec.~3.1 in the textbook~\cite{Poisson} for the proof.)
Here the minus sign is chosen so that $k^\mu$ is future-directed when $\Phi$ increases toward the future.
We install intrinsic coordinates $y^a=(\lambda, \theta^A)~(A=2,3,\cdots,n-1)$ on $\Sigma$, where $\lambda$ is an arbitrary parameter on the null generators of $\Sigma$ and the other $n-2$ coordinates $\theta^A$ label the generators. 
$\lambda$ can be an affine parameter on one side of $\Sigma$ but it is in general not possible on both sides of $\Sigma$.

The tangent vectors $e^\mu_{a}:=\partial x^\mu/\partial y^a$ on each side of $\Sigma$ are naturally segregated into a null vector $k^\mu$ that is tangent to the generators and spacelike vectors $e^\mu_{A}$ that point in the directions transverse to the generators.
$k^\mu$ and $e^\mu_{A}$ are explicitly written as
\begin{eqnarray}
k^\mu\equiv e^\mu_{\lambda}=\biggl(\frac{\partial x^\mu}{\partial \lambda}\biggl)_{\theta^A},~~~~e^\mu_{A}=\biggl(\frac{\partial x^\mu}{\partial \theta^A}\biggl)_{\lambda},
\end{eqnarray}
which satisfy $k^\mu k_{\mu}=0=k_{\mu}e^\mu_{A}$.
The line element on $\Sigma$ is
\begin{align}
&\D s_\Sigma^2=g_{\mu\nu}e^\mu_ae^\nu_b\D y^a\D y^b=\sigma_{AB}\D \theta^A \D \theta^B, \label{line-null}
\end{align}
where the induced metric $\sigma_{AB}$ on $\Sigma$ is defined by 
\begin{align}
&\sigma_{AB}:=g_{\mu\nu}e^\mu_Ae^\nu_B.\label{sigma-null}
\end{align}
A basis is completed by adding a transverse null vector $N^\mu$ which satisfies
\begin{eqnarray}
N_\mu N^{\mu}=0,~~~~N_\mu k^{\mu}=-1,~~~~N_{\mu}e^\mu_{A}=0. \label{complete}
\end{eqnarray}
The completeness relations of the basis are given as
\begin{eqnarray}
g^{\mu\nu}=-k^\mu N^\nu-N^\mu k^\nu+\sigma^{AB} e^\mu_{A}e^\nu_{B}, \label{comp}
\end{eqnarray}
where $\sigma^{AB}$ is the inverse of $\sigma_{AB}$.
The first junction conditions at $\Sigma$ are given by 
\begin{align}
[\sigma_{AB}]=0, \label{first-null}
\end{align}
which means that the induced metric on $\Sigma$ is the same on both sides of $\Sigma$.

Since $k^\mu$ is not normal but tangent to the generators of $\Sigma$, we introduce the transverse curvature $C_{ab}$ that properly represents the transverse derivative of the metric:
\begin{align}
C_{ab}&:=\frac12 ({\cal L}_N g_{\mu\nu})e^\mu_a e^\nu_b=(\nabla_\mu N_{\nu})e^\mu_{a}e^\nu_b, \label{Cab-def-original}
\end{align}
where we have used that $\nabla_\nu(N_\mu e^\mu_a)=0$ and an identity $(\nabla_\nu e^\mu_{a})e^\nu_b\equiv (\nabla_\nu e^\mu_{b})e^\nu_{a}$ at the last equality.
The jump in the transverse curvature $[C_{ab}]$ is directly related to the induced energy-momentum tensor $t_{\mu \nu}$ on $\Sigma$.

In the reformulation by Poisson~\cite{Poisson:2002nv}, one needs to introduce an arbitrary congruence of timelike geodesics $\gamma$ that arbitrarily intersect $\Sigma$, of which unit tangent vector is $u^\mu$, in order to derive the expression of $t_{\mu \nu}$. 
Each member of the congruence corresponds to the world line of a geodesic observer that intersects $\Sigma$ and performs measurements there.
The congruence corresponds to a whole family of such observers and gives operational meaning to the distributional character of $t_{\mu\nu}$.
We parametrize $\gamma$ by the proper time $\tau$ such that $\tau=0$ at $\Sigma$, $\tau<0$ in ${\cal M}_-$, and $\tau>0$ in ${\cal M}_+$.
Then, a displacement along a member of the congruence is described by $\D x^\mu=u^\mu\D \tau$.
Continuity of $u^\mu$ across $\Sigma$ requires
\begin{align}
[-u_\mu k^\mu]=0, \qquad [u_\mu e^\mu_A]=0,
\end{align}
while $u_\mu N^\mu$ may be discontinuous across $\Sigma$.

Then, under the first junction conditions (\ref{first-null}), one obtains the second junction conditions from the gravitational field equations, which determine the induced energy-momentum tensor $t_{\mu \nu}$ on $\Sigma$.
Generally, $t_{\mu \nu}$ is obtained in the following form:
\begin{align}
t_{\mu \nu}= (-k_{\eta}u^\eta)^{-1}S_{\mu\nu}, \label{null-t-J1}
\end{align}
where
\begin{align}
S_{\mu \nu}:=\mu k_\mu k_\nu +j_A(k_\mu e_{\nu}^A+e_{\mu}^A k_\nu)+p\sigma_{AB}e_\mu^A e_\nu^B.\label{null-S-J1}
\end{align}
Here $\mu$, $j^A$, and $p$ are respectively interpreted as the shell's surface density, surface current, and isotropic surface pressure.
Those quantities multiplied by $(-k_{\eta}u^\eta)^{-1}$ are the quantities that the geodesic observer corresponding to $\gamma$ measures.
$t_{\mu\nu}$ is symmetric and tangent to $\Sigma$, so that $t_{\mu \nu}k^\nu=0$ holds.
In general relativity, the second junction conditions are obtained from the Einstein equations $G_{\mu\nu}+\Lambda g_{\mu\nu}=\kappa_nT_{\mu\nu}$ and referred to as the {\it Barrab\`{e}s-Israel junction conditions}~\cite{Barrabes:1991ng,Poisson:2002nv}, which are given by 
\begin{align}
\kappa_n\mu=-\sigma^{AB}[C_{AB}],\qquad \kappa_nj^A=\sigma^{AB}[C_{\lambda B}],\qquad \kappa_np=-[C_{\lambda\lambda}]. \label{components3-n2}
\end{align}

\subsection{Main results}
\label{sec:main}


Now we are ready to present our main results.
If $\Sigma$ is timelike ($\varepsilon=1$), $h_{ab}$ has the Lorentzian signature and consequently the induced energy-momentum tensor $t^{\mu \nu}(=t^{ab}e^\mu_a e^\nu_b)$ can be any of the Hawking-Ellis types I through IV.
In contrast, on a spacelike $\Sigma$ ($\varepsilon=-1$), as shown below, $t^{\mu \nu}$ is of type I and simple equivalent representations of the standard energy conditions for $t^{\mu \nu}$ are available.
\begin{Prop}
\label{Prop:EC-spacelike}
An induced energy-momentum tensor $t_{\mu\nu}$ on a spacelike hypersurface $\Sigma$ is of the Hawking-Ellis type I and a non-vanishing $t_{\mu\nu}$ violates the DEC.
The NEC, WEC, and SEC are all equivalent to that $p_i\ge 0$ are satisfied for all $i(=1,2,\cdots,n-1)$, where $p_i$ are eigenvalues of the eigenvalue equations $t_{ab}v^{b}=\lambda h_{ab}v^{b}$.
\end{Prop}
{\it Proof:}
For $\varepsilon=-1$, we can choose $E^\mu_{(0)}=n^\mu$ and set $t_{(a)(b)}=t_{\mu\nu}E^\mu_{(a)}E^\nu_{(b)}$ ($a,b=1,2,\cdots,n-1$) be diagonal without loss of generality by using degrees of freedom to rotate the spacelike basis vectors $E^\mu_{(a)}$.
In this orthonormal frame, $t_{(\alpha)(\beta)}$ is given in the form of the Hawking-Ellis type I as $t_{(\alpha)(\beta)}=\mbox{diag}(0,p_1,p_2,\cdots,p_{n-1})$ and the proposition follows from Eqs.~(\ref{NEC-I})--(\ref{SEC-I}) because $p_i$ are eigenvalues of the eigenvalue equations $t_{(a)(b)}v^{(b)}=\lambda \delta_{(a)(b)}v^{(b)}$.
By identifications $E^\mu_{(a)}\equiv e^\mu_a/|e^\mu_a|$, the eigenvalue equations are written as $t_{ab}v^{b}=\lambda h_{ab}v^{b}$, where $v^b=v^\mu e^b_\mu$ is a vector on $\Sigma$.
\qed

Next, we consider the case where $\Sigma$ is null.
The Hawking-Ellis type of $t_{\mu\nu}$ and equivalent representations of the standard energy conditions for $t_{\mu\nu}$ are given as follows.
\begin{Prop}
\label{Prop:EC-null}
Define $J^2$ for an induced energy-momentum tensor $t_{\mu\nu}$ in the most general form~(\ref{null-t-J1}) on a null hypersurface $\Sigma$ by 
\begin{align}
J^2:=j_A j_B \sigma^{AB}. \label{def-J}
\end{align}
For $t_{\mu\nu}$, the Hawking-Ellis type and equivalent representations of the standard energy conditions are as shown in the following table.
\begin{center}
\begin{tabular}{|c|c||c|c|}
\hline \hline
& {\rm Hawking-Ellis type} & {\rm NEC, WEC, SEC} & {\rm DEC} \\\hline
$J=0$, $\mu=0$ & {\rm I} & $p\ge 0$ & $p=0$ \\ \hline
$J=0$, $\mu\ne 0$ & {\rm II} & $\mu> 0$, $p\ge 0$ & $\mu> 0$, $p=0$ \\ \hline
$J\ne 0$, $p=0$ & {\rm III} & {\rm violated} & {\rm violated} \\ \hline
$Jp\ne 0$, $J^2\ne \mu p$ & {\rm II} &$\mu p>J^2$, $p> 0$ & {\rm violated} \\ \hline
$Jp\ne 0$, $J^2= \mu p$ & {\rm I} & $p>0$ & {\rm violated} \\ 
\hline \hline
\end{tabular} 
\end{center} 
\end{Prop}
{\it Proof}. 
We introduce orthonormal basis vectors $E^\mu_{(\alpha)}$ at the location of $\Sigma$ such that a timelike basis vector $E^\mu_{(0)}$ and a spacelike basis vector $E^\mu_{(1)}$ are given by 
\begin{align}
\left\{
\begin{array}{ll}
E^\mu_{(0)}=(k^\mu+N^\mu)/\sqrt{2}\\
E^\mu_{(1)}=(-k^\mu+N^\mu)/\sqrt{2}
\end{array}
\right.~~\Leftrightarrow~~
\left\{
\begin{array}{ll}
k^\mu=(E^\mu_{(0)}-E^\mu_{(1)})/\sqrt{2}\\
N^\mu=(E^\mu_{(0)}+E^\mu_{(1)})/\sqrt{2}
\end{array}
\right..
\end{align}
If $j_A e_{\mu}^A$ is non-vanishing, using degrees of freedom to rotate the spacelike basis vectors $E^\mu_{(i)}$ ($i=2,3,\cdots,n-1$), we can set $E_{\mu(2)}$ point the direction of $j_A e_{\mu}^A$ such that $j_A e_{\mu}^A=-JE_{\mu(2)}$ without loss of generality, where $J$ satisfies Eq.~(\ref{def-J}).
If $j_A e_{\mu}^A$ is vanishing (and then $J^2=0$), we don't specify the direction of $E_{\mu(2)}$.
Now Eq.~(\ref{null-S-J1}) is written as
\begin{align}
S^{\mu \nu}=&\frac{1}{2}\mu(E^\mu_{(0)}-E^\mu_{(1)})(E^\nu_{(0)}-E^\nu_{(1)}) -\frac{1}{\sqrt{2}}J\left[(E^\mu_{(0)}-E^\mu_{(1)})E^\nu_{(2)}+E^\mu_{(2)}(E^\nu_{(0)}-E^\nu_{(1)})\right] \nonumber\\
&+p\left[g^{\mu\nu}+\frac{1}{2}(E^\mu_{(0)}-E^\mu_{(1)})(E^\nu_{(0)}+E^\nu_{(1)}) +\frac{1}{2}(E^\mu_{(0)}+E^\mu_{(1)})(E^\nu_{(0)}-E^\nu_{(1)}) \right],
\end{align}
where we have used Eq.~(\ref{comp}).
Then, we obtain orthonormal components of the induced energy-momentum tensor (\ref{null-t-J1}) as $t^{(\alpha)(\beta)}=(-k_{\eta}u^\eta)^{-1}S^{(\alpha)(\beta)}$, where 
\begin{equation} 
\label{S-large}
S^{(\alpha)(\beta)}:=\left( 
\vphantom{\begin{array}{c}1\\1\\1\\1\\1\\1\\1\end{array}}
\begin{array}{ccccccc}
{\mu}/2 &{\mu}/2&{J}/\sqrt{2}&0&0&\cdots &0\\
{\mu}/2 &{\mu}/2&{J}/\sqrt{2}&0&0&\cdots &0\\
{J}/\sqrt{2}&{J}/\sqrt{2}&{p}&0&0&\cdots &0\\
0&0&0&{p}&0&\cdots&0 \\
0&0&0&0&\ddots&\vdots&\vdots \\
\vdots&\vdots&\vdots&\vdots&\cdots&\ddots&0\\
0&0&0&0 &\cdots&0&{p}
\end{array}
\right).
\end{equation}
Since the factor $(-k_{\eta}u^\eta)^{-1}$ is positive, the Hawking-Ellis types and the energy conditions for ${t}^{(\alpha)(\beta)}$ and $S^{(\alpha)(\beta)}$ are the same.

For  $J=0$, $S^{(\alpha)(\beta)}$ is in the canonical form~(\ref{T-typeII}) with $\nu=\mu/2$, $\rho=0$, and $p_2=p_3=\cdots=p_{n-1}=p$. 
Specifically, it reduces to the type I form if $\mu = 0$ and to the type II form if $\mu \neq 0$.
Then, the standard energy conditions are equivalent to
\begin{align}
\mbox{NEC,~WEC,~SEC}:&~~\mu\ge 0\mbox{~and~}p\ge 0,\label{NEC-II-J=0}\\
\mbox{DEC}:&~~\mu\ge 0\mbox{~and~}p=0 \label{DEC-II-J=0}
\end{align}
by Eqs.~(\ref{NEC-I})--(\ref{SEC-I}) and Eqs.~(\ref{NEC-II})--(\ref{SEC-II}).
For $J\ne 0$ with $p=0$, $S^{(\alpha)(\beta)}$ is in the canonical type-III form~(\ref{T-typeIII}) with $\nu=\mu/2$, $\rho=0$, and $\zeta= J/\sqrt{2}$, and then all the standard energy conditions are violated.
Hereafter we assume $Jp\ne 0$.

The characteristic equation of the eigenvalue equations gives $\lambda^2(\lambda-p)^{n-2}=0$ and hence the eigenvalues are $\lambda=\{0,p\}$.
For the eigenvalue $\lambda=p(\ne 0)$, the corresponding $n-2$ Lorentz-covariant eigenvectors are spacelike and their normalized forms are given by $n_{(\alpha)}=\{w_{i(\alpha)}\}$ ($i=2,3,\cdots,n-1$), where
\begin{equation} 
\label{eigen-vec-null}
w_{2(\alpha)}=\left( 
\vphantom{\begin{array}{c}1\\1\\1\\1\\1\end{array}}
\begin{array}{c}
J/(\sqrt{2}p)\\
-J/(\sqrt{2}p)\\
-1\\
0\\
\vdots \\
0
\end{array}
\right),\quad
w_{3(\alpha)}=\left( 
\vphantom{\begin{array}{c}1\\1\\1\\1\\1\end{array}}
\begin{array}{c}
0\\
0\\
0\\
1\\
\vdots \\
0
\end{array}
\right),\quad \cdots,\quad
w_{n-1(\alpha)}=\left( 
\vphantom{\begin{array}{c}1\\1\\1\\1\\1\end{array}}
\begin{array}{c}
0\\
0\\
0\\
0\\
\vdots \\
1
\end{array}
\right).
\end{equation}
Comparing 
\begin{align} 
&S^{(\alpha)(\beta)}-\sum_{i=2}^{n-1}pw_{i}^{(\alpha)} w_i^{(\beta)} \nonumber\\
&~~~~~~~~~~=\left( 
\vphantom{\begin{array}{c}1\\1\\1\\1\\1\\1\\1\end{array}}
\begin{array}{ccccccc}
(\mu p-J^2)/(2p) &(\mu p-J^2)/(2p)&0&0&0&\cdots &0\\
(\mu p-J^2)/(2p) &(\mu p-J^2)/(2p)&0&0&0&\cdots &0\\
0&0&0&0&0&\cdots &0\\
0&0&0&{0}&0&\cdots&0 \\
0&0&0&0&\ddots&\vdots&\vdots \\
\vdots&\vdots&\vdots&\vdots&\cdots&\ddots&0\\
0&0&0&0 &\cdots&0&{0}
\end{array}
\right)
\end{align}
with Eqs.~(\ref{T-typeII}) and (\ref{typeII-decomp}), we find that $S^{(\alpha)(\beta)}$ is of type II for $\mu p\ne J^2$ and of type I for $\mu p=J^2$.
For $\mu p\ne J^2$, the canonical form of $S^{(\alpha)(\beta)}$ is given by Eq.~(\ref{T-typeII}) with $\nu=(\mu p-J^2)/(2p)$, $\rho=0$, and $p_2=p_3=\cdots=p_{n-1}=p$.
Then, by Eqs.~(\ref{NEC-II})--(\ref{SEC-II}), the NEC, WEC, and SEC are equivalent to $\mu p>J^2$ and $p>0$, while the DEC is violated.
For $\mu p= J^2$, the canonical form of $S^{(\alpha)(\beta)}$ is given by Eq.~(\ref{T-typeI}) with $\rho=0$, $p_1=0$, and $p_2=p_3=\cdots=p_{n-1}=p$.
Then, by Eqs.~(\ref{NEC-I})--(\ref{SEC-I}), the NEC, WEC, and SEC are equivalent to $p>0$, while the DEC is violated.
\qed

Propositions~\ref{Prop:EC-spacelike} and \ref{Prop:EC-null} are the main results in the present paper.
Without imposing a spacetime symmetry, we have shown that the induced energy-momentum tensor $t_{\mu\nu}$ on a shell $\Sigma$ is of the Hawking-Ellis type I if $\Sigma$ is spacelike and either of type I, II, or III if $\Sigma$ is null.
Then, we have derived equivalent representations of the standard energy conditions for $t_{\mu\nu}$.
In particular, on a spacelike shell or on a null shell with non-vanishing surface current, $t_{\mu\nu}$ inevitably violates the DEC.
If the surface pressure on the null shell is vanishing in addition, $t_{\mu\nu}$ is of type III and violates all the standard energy conditions.
Those fully general results can be used in {\it any} theory of gravity.

Lastly, it is emphasized that Proposition~\ref{Prop:EC-null} can be applied not only to an induced energy-momentum tensor on a null shell but also to {\it any} energy-momentum tensor given in the following form
\begin{align}
T_{\mu \nu}=\mu k_\mu k_\nu +j_A(k_\mu e_{\nu}^A+e_{\mu}^A k_\nu)+p\sigma_{AB}e_\mu^A e_\nu^B.\label{null-matter}
\end{align}
For example, the above energy-momentum tensor with $j_A\equiv 0$ and $p\equiv 0$ corresponds to a null dust fluid, which is of the Hawking-Ellis type II. (See Sec.~4.2 in~\cite{Maeda:2018hqu}.)
Also, a matter field described by the energy-momentum tensor (\ref{null-matter}) with $p\equiv 0$ is referred to as a {\it gyraton}, which is of the Hawking-Ellis type III. (See~\cite{Podolsky:2018zha} for $n=3$ and~\cite{Martin-Moruno:2019kzc} for $n=4$.)

\section{Applications in general relativity in four dimensions}
\label{sec:apply}

In this section, as a demonstration, we apply Propositions~\ref{Prop:EC-spacelike} and \ref{Prop:EC-null} to several physical situations in general relativity in four dimensions ($n=4$).

\subsection{Shells in the Schwarzschild spacetime}

\subsubsection{Black bounce with a spacelike shell}

\begin{figure}[htbp]
\begin{center}
\includegraphics[width=1.0\linewidth]{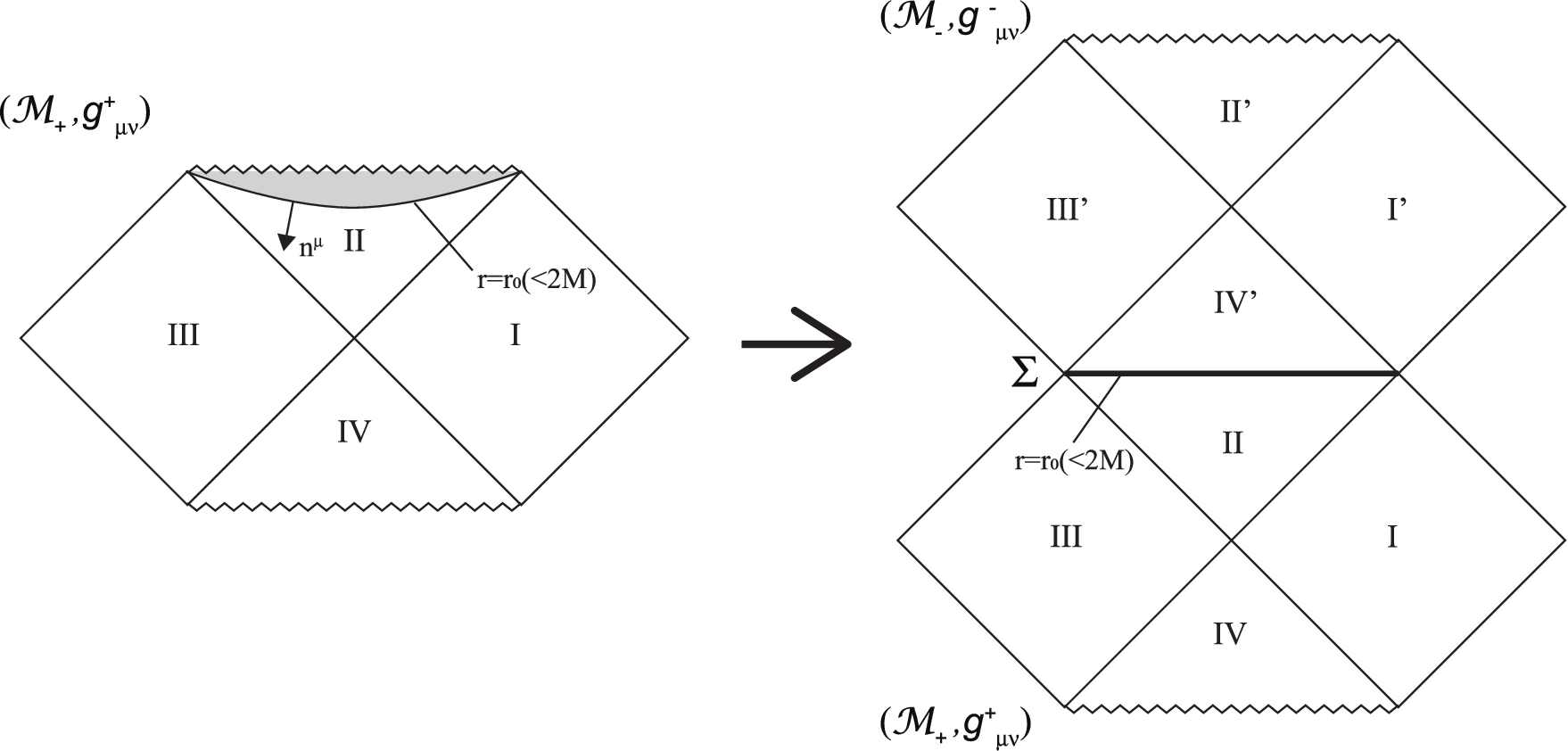}
\caption{\label{Fig-Schwarzschild-spacelike} A Penrose diagram of the thin-shell black-bounce spacetime constructed by gluing two identical Schwarzschild spacetimes at a spacelike hypersurface $r=r_0(<2M)$.}
\end{center}
\end{figure}

As the first application, we consider a spacelike massive thin shell constructed by gluing two Schwarzschild bulk spacetimes.
We write the bulk metric in the diagonal coordinates $x^\mu=(t,r,\theta,\phi)$ as
\begin{align}
&\D s^2=g_{\mu\nu}\D x^\mu \D x^\nu=-f(r)\D t^2+f(r)^{-1}\D r^2+r^2\gamma_{AB}\D z^A\D z^B, \label{Lambda-vacuum}\\
&f(r) := 1-\frac{2M}{r},\qquad \gamma_{AB}\D z^A\D z^B=\D\theta^2+\sin^2\theta\D\phi^2,\label{f-def}
\end{align}
where $M$ is a positive mass parameter and $z^A (A=2,3)$ are coordinates on the unit two-sphere ${\rm S}^2$.
Let $r_0$ be a constant satisfying $0<r_0<2M$ and consider two spacetimes which are described by the metric~(\ref{Lambda-vacuum}) with the same positive mass $M$ and defined in the domain $r\ge r_0$.
We glue them at a spacelike hypersurface $\Sigma$ described by $r=r_0$.
In the resulting spacetime, the big bounce occurs at a spacelike bounce hypersurface $\Sigma$ inside the event horizon of a black hole as shown in Fig.~\ref{Fig-Schwarzschild-spacelike}.
Such a spacetime is referred to as a {\it black bounce}.
Our model is a thin-shell version of the Simpson-Visser black-bounce model~\cite{Simpson:2018tsi}, in which the metric is analytic everywhere.
We will show that all the standard energy conditions are violated on $\Sigma$.

As seen from ${\cal M}_+$, the unit normal one-form to $\Sigma$ is given by 
\begin{equation}
n^+_\mu \D x^\mu =-\frac{1}{\sqrt{-f(r_0)}}\D r\, .\label{n-def-s}
\end{equation}
Note that the timelike vector $n^\mu=(0,\sqrt{-f(r_0)},0,0)$ points an increasing direction of $r$, which is consistent with the assumption in Sec.~\ref{sec:non-null} that $n^\mu$ points from ${\cal M}_-$ to ${\cal M}_+$. 
The induced metric $h_{ab}$ on the spacelike $\Sigma$ is given by 
\begin{equation}
\label{1stout} 
\D s_\Sigma^2=h_{ab}(y)\D y^a\D y^b= -f(r_0)\D t^2+r_0^2\gamma_{AB}\D z^A\D z^B\, ,
\end{equation}
where $a=1,2,3$ and we have identified $y^1\equiv t$ and $y^A\equiv z^A$.
Since $r_0=2M$ is the same on both sides of $\Sigma$, the first junction conditions $[h_{ab}]=0$ are satisfied.

Using Eq.~(\ref{K-def}) with Eq.~(\ref{n-def-s}) and 
\begin{equation}
e^\mu_t\frac{\partial}{\partial x^\mu}=\frac{\partial}{\partial t},\qquad e^\mu_A\frac{\partial}{\partial x^\mu}=\frac{\partial}{\partial z^A},
\end{equation}
we obtain non-zero components of $K_{ab}$ as seen from ${\cal M}_+$ as
\begin{align}
K^+_{tt}=&-\frac12\sqrt{-f}f'|_{r=r_0}\,, \quad K^+_{AB}=r\sqrt{-f}\gamma _{AB}|_{r=r_0}\, ,
\end{align}
which give
\begin{align}
K^+=K^+_{ab}h^{ab}=\biggl(-\frac{f'}{2\sqrt{-f}}+\frac{2\sqrt{-f}}{r}\biggl)\biggl|_{r=r_0}\, ,
\end{align}
where a prime denotes differentiation with respect to $r$.
As seen from ${\cal M}_-$, the unit normal is given by 
\begin{equation}
n^-_\mu \D x^\mu =\frac{1}{\sqrt{-f(r_0)}}\D r\, 
\end{equation}
instead of Eq.~(\ref{n-def-s}), which points a decreasing direction of $r$.
As a result, $K_{ab}$ as seen from ${\cal M}_-$ is given by $K_{ab}^-=-K_{ab}^+$.

Then, the Israel junction conditions (\ref{j-eq-summary1}) with $\varepsilon=-1$ give
\begin{equation} 
\kappa_4t_{ab}=2(K_{ab}^+-h_{ab}K^+),\label{tab-K}
\end{equation}
of which non-zero components are given by
\begin{align}
t_{tt}=&-\frac{4(-f)^{3/2}}{\kappa_4r}\biggl|_{r=r_0}=:\lambda_1,\qquad t_{AB}=\frac{r(rf'+2f)}{\kappa_4\sqrt{-f}}\gamma_{AB}\biggl|_{r=r_0}=:\lambda_2\gamma_{AB}.\label{t_AB-s}
\end{align}
Eigenvalues of $t_{ab}v^{b}=\lambda h_{ab}v^{b}$ are given by $\lambda=\{p_1,p_2\}$, where
\begin{align}
p_1:=-\frac{4\sqrt{-f}}{\kappa_4r}\biggl|_{r=r_0}, \qquad p_2:=\frac{rf'+2f}{\kappa_4r\sqrt{-f}}\biggl|_{r=r_0}.
\end{align}
Since $p_1$ is negative, all the standard energy conditions are violated on $\Sigma$ by Proposition~\ref{Prop:EC-spacelike}.

\subsubsection{Lightlike impulse as a null shell}

\begin{figure}[htbp]
\begin{center}
\includegraphics[width=0.7\linewidth]{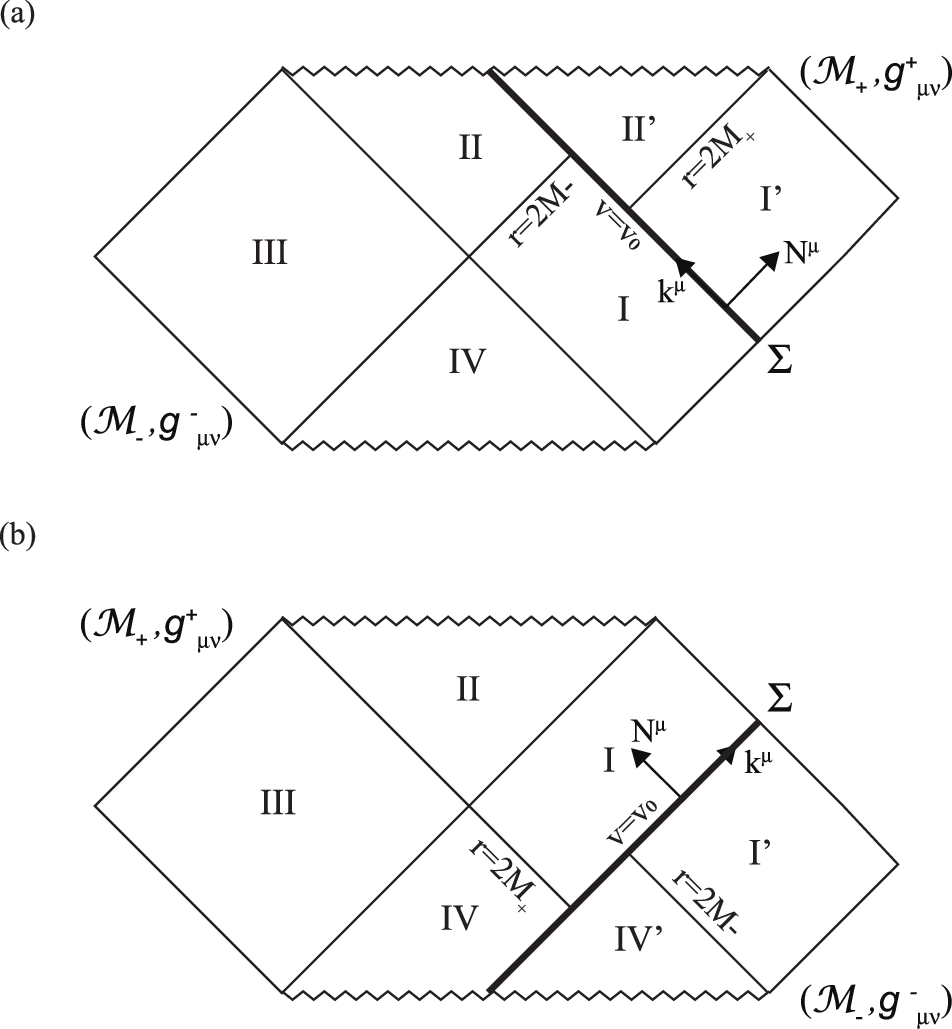}
\caption{\label{Fig-Schwarzschild-null} Penrose diagrams of the Schwarzschild spacetime with a lightlike impulse as a null shell for (a) $\epsilon=1$ with $M_+>M_-$ and (b) $\epsilon=-1$ with $M_+<M_-$. These shells satisfy all the standard energy conditions.}
\end{center}
\end{figure}

Next, we study the energy conditions of a radially propagating lightlike impulse in the Schwarzschild spacetime described by a null shell, which has been studied in~\cite{Barrabes:1991ng,Poisson:2002nv}.
We write the bulk metric in the single-null coordinates $(v,r,z^A)$ as
\begin{align}
\label{bulk-null}
\D s^2=&g_{\mu\nu}\D x^\mu \D x^\nu=-f(r)\D v^2+2\epsilon \D v\D r+r^2\gamma_{AB}\D z^A\D z^B,
\end{align}
where $f(r)$ and $\gamma_{AB}$ are defined by Eq.~(\ref{f-def}).
Here $\epsilon=1(-1)$ is chosen to construct a collapsing (spreading) shell.
Non-zero components of the inverse metric are given by
\begin{align}
g^{vv}=0,\qquad g^{vr}=\epsilon,\qquad g^{rr}=f,\qquad g^{AB}=r^{-2}\gamma^{AB},
\end{align}
where $\gamma^{AB}$ is the inverse of $\gamma_{AB}$.
For a given $\epsilon$, we consider a spacetime described by the metric~(\ref{bulk-null}) with $M=M_+(M_-)$ defined in the domain $v\ge(\le) v_0$ as ${\cal M}_+({\cal M}_-)$, where $v_0$ is a constant.
We glue them at a null hypersurface $\Sigma$ defined by $v=v_0$ and the resulting spacetime describes a radially collapsing (for $\epsilon=1$) or spreading (for $\epsilon=-1$) lightlike impulse in the Schwarzschild spacetime as shown in Fig.~\ref{Fig-Schwarzschild-null}.
We will derive the equivalent inequalities to the standard energy conditions on $\Sigma$.

The null vector $k^\mu=-g^{\mu\nu}\nabla_\nu v$ tangent to the null generators of $\Sigma$ is given by
\begin{align}
k^\mu\frac{\partial}{\partial x^\mu}=-\epsilon\frac{\partial}{\partial r},
\end{align}
which satisfies $k_\mu k^\mu=0$ and $k^\nu\nabla_\nu k^\mu=0$.
Hence, $k^\mu$ is parametrized by an affine parameter $\lambda$. 
By $k^r=\D r/\D\lambda=-\epsilon$, we identify $-\epsilon r$ with ${\lambda}$.
Since $z^A$ are constant along the generators, we install coordinates $y^a=({\lambda},\theta^A)$ on $\Sigma$ such that $\lambda=-\epsilon r$ and $\theta^A=z^A$.
Now the parametric equations $x^\mu=x^\mu({\lambda},\theta^A)$ describing $\Sigma$ are $v=v_0$, $r=-\epsilon {\lambda}$, and $z^A=\theta^A$, where $\theta^A$ label the generators of $\Sigma$.
$e^\mu_a := \partial x^\mu/\partial y^a$ are given by
\begin{align}
e^\mu_{\lambda}\frac{\partial}{\partial x^\mu}=k^\mu\frac{\partial}{\partial x^\mu},\qquad e^\mu_A\frac{\partial}{\partial x^\mu}=\frac{\partial}{\partial z^A}.
\end{align}
By Eq.~(\ref{line-null}), the induced metric on $\Sigma$ is given by
\begin{align}
\D s_{\Sigma}^2= \sigma_{AB}\D \theta^A \D \theta^B={\lambda}^2\gamma_{AB}\D z^A\D z^B. \label{hab}
\end{align}
Since $\sigma_{AB}$ is independent from $M_\pm $, the first junction conditions $[\sigma_{AB}]=0$ are satisfied.

The transverse null vector $N^\mu$ completing the basis is given by
\begin{align}
N^\mu \frac{\partial}{\partial x^\mu}=\frac{\partial}{\partial 
v}+\frac12\epsilon f\frac{\partial}{\partial r}.\label{N-attachment}
\end{align}
The expression $N_\mu \D x^\mu=-(f/2)\D v+\epsilon \D r$ shows $N_\mu N^\mu 
=0$, $N_\mu e^\mu_{\lambda}=-1$, and $N_\mu e^\mu_A=0$.
Then, by Eq.~(\ref{Cab-def-original}), nonvanishing components of the transverse curvature of $\Sigma$ are computed to give
\begin{align}
C_{AB}=\frac12\epsilon rf\gamma_{AB}\biggl|_{r=-\epsilon {\lambda}}.\label{trans-C0}
\end{align}
Now we have $[C_{\lambda\lambda}]=[C_{\lambda A}]=0$ and
\begin{align}
[C_{AB}]=&\frac12\epsilon r(f_+-f_-)\gamma_{AB}\biggl|_{r=-\epsilon {\lambda}}=-\epsilon (M_+-M_-)\gamma_{AB}.
\end{align}
Then, the Barrab\`{e}s-Israel junction conditions~(\ref{components3-n2}) give $p=0$, $j^A=0$, and 
\begin{align}
\mu=\frac{2\epsilon(M_+-M_-)}{\kappa_4\lambda^2}.
\end{align}
As $p=0$ and $J=0$ are satisfied, $t^{\mu\nu}$ on $\Sigma$ is of the Hawking-Ellis type II and all the standard energy conditions are satisfied (violated) for $\epsilon(M_+-M_-)\ge(<) 0$ by Proposition~\ref{Prop:EC-null}.

\subsubsection{Accretion of a slowly-rotating null shell}

As the third application, we consider a situation that a slowly-rotating null shell is collapsing in the Schwarzschild spacetime with its mass $M-m$.
This system has been studied in~\cite{Poisson:2002nv}.
The past (or interior) spacetime ${\cal M}_-$ of the shell is described by the Schwarzschild spacetime metric
\begin{align}
&\D s_-^2=-F(r)\D {\bar t}^2+F(r)^{-1}\D r^2+r^2(\D\theta^2+\sin^2\theta\D \varphi^2),\\
&F(r):=1-\frac{2(M-m)}{r}
\end{align}
in the coordinates $x_-^\mu=({\bar t},r,\theta,\varphi)$, where $M$ and $m$ are constant.
The future (or exterior) spacetime ${\cal M}_+$ is described by the slowly rotating Kerr metric:
\begin{align}
\label{slow-Kerr}
\begin{aligned}
&\D s_+^2=-f(r)\D {t}^2+f(r)^{-1}\D r^2+r^2(\D\theta^2+\sin^2\theta\D \phi^2)-\frac{4Ma}{r}\sin^2\theta\D t\D\phi,\\
&f(r):=1-\frac{2M}{r}
\end{aligned}
\end{align}
in the coordinates $x_+^\mu=(t,r,\theta,\phi)$, where $a$ is a rotation parameter.
Non-zero components of the inverse metric on ${\cal M}_+$ are 
\begin{align}
&g^{tt}=-\frac{r^4}{r^4f+4M^2a^2\sin^2\theta},\qquad g^{t\phi}=-\frac{2Mar}{r^4f+4M^2a^2\sin^2\theta},\\
&g^{rr}=f,\qquad g^{\theta\theta}=r^{-2},\qquad g^{\phi\phi}=\frac{r^2f}{\sin^2\theta(r^4f+4M^2a^2\sin^2\theta)}.
\end{align}
The metric (\ref{slow-Kerr}) is a solution to the vacuum Einstein equations in the slow-rotation approximation, in which we consider up to the linear order of $a$.
We will show that, due to the ambiguity to define a null vector, one cannot obtain a definite conclusion on the Hawking-Ellis type of the shell and the energy conditions for the null shell in this approximation.

On ${\cal M}_+$, we define $v=v(t,r)$ and $r_*=r_*(r)$ by 
\begin{align}
&v:=t+r_*,\\
&r_*:=r+2M\ln\biggl|\frac{r}{2M}-1\biggl|\biggl(=\int f(r)^{-1}\D r\biggl)
\end{align}
and let $\Sigma$ be a hypersurface described by $v=v_0=$constant.
A vector $k^\mu=-g^{\mu\nu}\nabla_\nu v$ is given by 
\begin{align}
k^\mu\frac{\partial}{\partial x^\mu}=&-g^{tt}\frac{\partial}{\partial t}-\frac{\partial}{\partial r}-g^{t\phi}\frac{\partial}{\partial \phi} \nonumber\\
\simeq& f^{-1}\frac{\partial}{\partial t}-\frac{\partial}{\partial r}+\frac{2Ma}{r^3f}\frac{\partial}{\partial \phi},
\end{align}
which satisfies
\begin{align}
k_\mu k^\mu=&\frac{4M^2a^2\sin^2\theta}{f(r^4f+4M^2a^2\sin^2\theta)}\simeq 0,\\
k^\nu\nabla_\nu k^\mu\frac{\partial}{\partial x^\mu}=&-\frac{4M^2a^2\sin^2\theta[2r^3f^2+f'(r^4f+2M^2a^2\sin^2\theta)]}{f(r^4f+4M^2a^2\sin^2\theta)^2}\frac{\partial}{\partial r} \nonumber \\
&+\frac{4M^2a^2r^2\sin\theta\cos\theta}{(r^4f+4M^2a^2\sin^2\theta)^2}\frac{\partial}{\partial \theta}\simeq 0.
\end{align}
Hence, $k^\mu$ is null and affinely parametrized in the slow-rotation approximation.
In this approximation, $\Sigma$ is a null hypersurface and $k^\mu$ is tangent to the null generators of $\Sigma$.
By $k^r=-1$ and $k^\theta=0$, the generators are affinely parametrized by $\lambda=-r$ and $\theta$ is constant on each generator.

On ${\cal M}_+$, we also define $\psi=\psi(r,\phi)$ by 
\begin{align}
\psi:=\phi+\frac{a}{r}\biggl(1+\frac{r}{2M}\ln f\biggl).\label{def-psi}
\end{align}
Along generators of $\Sigma$, we have
\begin{align}
&\frac{\D \phi}{\D r}=\frac{k^\phi}{k^r}\simeq-\frac{2Ma}{r^3f},\label{dphi/dr}
\end{align}
which is integrated to give
\begin{align}
\phi-\phi_0\simeq-\frac{a}{r}\biggl(1+\frac{r}{2M}\ln f\biggl),
\end{align}
where $\phi_0$ is an integration constant.
Therefore, $\psi$ defined by Eq.~(\ref{def-psi}) is constant on the generators of $\Sigma$ in the slow-rotation approximation.
Thus, we install coordinates $y^a=({\lambda},\theta^A)$ on $\Sigma$, where $\theta^A=(\theta,\psi)$.
Now the parametric equations $x^\mu=x^\mu({\lambda},\theta^A)$ describing $\Sigma$ as seen from ${\cal M}_+$ are 
\begin{align}
&t=-r_*(-\lambda)+v_0,\qquad r=-{\lambda}, \qquad \theta=\theta,\qquad \phi=\psi+\frac{a}{\lambda}\biggl(1-\frac{\lambda}{2M}\ln f(-\lambda)\biggl).
\end{align}
Then, $e^\mu_a := \partial x^\mu/\partial y^a$ are given by
\begin{align}
e^\mu_{\lambda}\frac{\partial}{\partial x^\mu}=k^\mu\frac{\partial}{\partial x^\mu},\qquad e^\mu_\theta\frac{\partial}{\partial x^\mu}=\frac{\partial}{\partial \theta},\qquad e^\mu_\psi\frac{\partial}{\partial x^\mu}=\frac{\partial}{\partial \phi},
\end{align}
which satisfy $k_\mu k^\mu\simeq 0$ and $k_\mu e^\mu_A=0$.
By Eq.~(\ref{line-null}), the induced metric on $\Sigma$ is given by
\begin{align}
\D s_{\Sigma}^2=\sigma_{AB}\D \theta^A \D \theta^B=\lambda^2(\D\theta^2+\sin^2\theta\D\psi^2).\label{hab-slow}
\end{align}
The transverse vector $N^\mu$ completing the basis in the slow-rotation approximation is given by
\begin{align}
N_\mu\D x^\mu=\frac12(-f\D t+\D r),
\end{align}
which satisfy $N_\mu k^\mu=-1$, $N_\mu e^\mu_A=0$, and 
\begin{align}
N_\mu N^\mu=\frac{fM^2a^2\sin^2\theta}{r^4f+4M^2a^2\sin^2\theta}\simeq 0.
\end{align}
From Eq.~(\ref{Cab-def-original}), the transverse curvature of $\Sigma$ is computed to give
\begin{align}
C^+_{\lambda\lambda}=&\frac{4M^2a^2\sin^2\theta[3r^3f^2+f'(r^4f+M^2a^2\sin^2\theta)]}{f(r^4f+4M^2a^2\sin^2\theta)^2}\biggl|_{r=-\lambda}\simeq 0,\\
C^+_{\lambda\theta}=&\frac{2M^2a^2r^4f \sin\theta\cos\theta}{(r^4f+4M^2a^2\sin^2\theta)^2}\biggl|_{r=-\lambda}\simeq 0,\\
C^+_{\lambda\psi}=&\frac{3Mar^2f\sin^2\theta}{r^4f+4M^2a^2\sin^2\theta}\biggl|_{r=-\lambda}\simeq \frac{3Ma\sin^2\theta}{r^2}\biggl|_{r=-\lambda},\\
C^+_{\theta\theta}=&\frac12rf\biggl|_{r=-\lambda},\qquad C^+_{\psi\psi}=\frac12rf\sin^2\theta\biggl|_{r=-\lambda}.
\end{align}

We can obtain the results on ${\cal M}_-$ by replacing such that $M\to M-m$, $a\to 0$, and $(t,\phi)\to ({\bar t},\varphi)$.
Thus, as seen from ${\cal M}_-$, we have $\D s_{\Sigma-}^2=\sigma_{AB}\D\theta^A\D\theta^B$ and 
\begin{align}
C^-_{\lambda\lambda}=&0,\qquad C^-_{\lambda\theta}=0,\qquad C^-_{\lambda\psi}=0,\qquad C^-_{AB}=\frac{F}{2r}\sigma_{AB}\biggl|_{r=-\lambda},
\end{align}
where $\sigma_{AB}$ is given by Eq.~(\ref{hab-slow}), so that the first junction conditions $[\sigma_{AB}]=0$ are satisfied.
Then, using the Barrab\`{e}s-Israel junction conditions (\ref{components3-n2}) and 
\begin{align}
[C_{\lambda\lambda}]= [C_{\lambda\theta}]=0,\qquad [C_{\lambda\psi}]=\frac{3Ma\sin^2\theta}{r^2}\biggl|_{r=-\lambda},\qquad [C_{AB}]=-\frac{m}{r^2}\sigma_{AB}\biggl|_{r=-\lambda},
\end{align}
we obtain
\begin{align}
\mu=&\frac{2m}{\kappa_4\lambda^2},\qquad j^\theta=0,\qquad j^\psi=\frac{3Ma}{\kappa_4\lambda^4},\qquad p=0.
\end{align}
Since we have $p=0$ and $J\ne 0$, the energy-momentum tensor $t_{\mu\nu}$ on the shell $\Sigma$ is of the Hawking-Ellis type III and violates all the standard energy conditions by Proposition~\ref{Prop:EC-null}.

However, this conclusion under the slow-rotation approximation cannot be definite as shown below.
Now $t_{\mu\nu}$ on the shell $\Sigma$ is given by 
\begin{align}
t^{\mu\nu}=&(-k_{\eta}u^\eta)^{-1}[\mu k^\mu k^\nu+j^\psi(k^\mu e^\nu_\psi+e^\mu_\psi k^\nu)].\label{t-III}
\end{align}
This $t^{\mu\nu}$ can be written in the slow-rotation approximation as 
\begin{align}
t^{\mu\nu}=&(-k_{\eta}u^\eta)^{-1}\mu \ell^\mu\ell^\nu, \label{t-II}\\
\ell^{\mu}:=&k^\mu+\frac{j^\psi}{\mu} e^\mu_\psi,
\end{align}
where components of $\ell^\mu$ in the coordinates $(t,r,\theta,\phi)$ are given by
\begin{align}
&\ell^{\mu}\simeq \biggl(f^{-1},-1,0,\frac{2Ma}{r^3f}+\frac{3Ma}{2mr^2}\biggl)\biggl|_{r=-\lambda}.
\end{align}
Since $\ell_\mu\ell^{\mu}\simeq 0$ is satisfied, $\ell^{\mu}$ is null in the slow-rotation approximation.
As a consequence, $t^{\mu\nu}$ in the form of Eq.~(\ref{t-II}) is of the Hawking-Ellis type II and all the standard energy conditions are satisfied (violated) for $\mu\ge (<)0$ by Proposition~\ref{Prop:EC-null} in spite that $t^{\mu\nu}$ in the form of Eq.~(\ref{t-III}) is of type III and all the standard energy conditions are inevitably violated.
Hence, due to the ambiguity to define a null vector, one cannot obtain a definite conclusion on the Hawking-Ellis type of $t_{\mu\nu}$ and the energy conditions on $\Sigma$.
To avoid this problem, higher-order effects of the rotation parameter $a$ must be taken into account.
We will see in the next subsection how the result in the full-order analysis is different from the one in the slow-rotation approximation in a cylindrically symmetric spacetime.

\subsection{Cylindrically symmetric rotating null shell}

As the fourth application, we consider a rotating cylindrically symmetric null shell collapsing in the Minkowski spacetime.
The past (or interior) spacetime ${\cal M}_-$ of the shell is described by the Minkowski metric
\begin{align}
\D s_-^2=-\D t^2+\D\rho^2+\rho^2\D\varphi^2+\D z^2 \label{flat}
\end{align}
in the cylindrical coordinates $x_-^\mu=(t,\rho,\varphi,z)$.
The domains of the coordinates are $t,z\in(-\infty,\infty)$, $\rho\in(0,\infty)$, and $\varphi\in[0,2\pi)$ and we identify $(t,\rho,\varphi,z)$ and $(t,\rho,\varphi+2\pi,z)$ so that $\rho=0$ is a coordinate singularity.
The future (or exterior) spacetime ${\cal M}_+$ of the shell is described by the following locally flat metric
\begin{align}
\D s_+^2=-(\D T+m\D \Phi)^2+{\cal C}^2\D r^2+r^2\D\Phi^2+\D z^2 \label{Lewis}
\end{align}
in the coordinates $x_+^\mu=(T,r,\Phi,z)$, where ${\cal C}(>0)$ and $m$ are constants.
The domains of the coordinates are $T,z\in(-\infty,\infty)$, $r\in(0,\infty)$, and $\Phi\in[0,2\pi)$ and we identify $(T,r,\Phi,z)$ and $(T,r,\Phi+2\pi,z)$ so that $r=0$ is a conical singularity.
In fact, the exterior metric (\ref{Lewis}) can be written as the interior (\ref{flat}) by coordinate transformations $t=T+m\Phi$, $\rho={\cal C}r$, and $\varphi={\cal C}^{-1}\Phi$, but still ${\cal M}_+$ is globally different from ${\cal M}_-$ due to the different identification $(t,\rho,\varphi,z)$ and $(t+2m\pi,\rho,\varphi+2\pi{\cal C}^{-1},z)$.

Non-vanishing components of the inverse metric on ${\cal M}_+$ are
\begin{align}
g^{TT}=-\frac{r^2-m^2}{r^2},\quad g^{T\psi}=-\frac{m}{r^2},\quad g^{rr}={\cal C}^{-2},\quad g^{\Phi\Phi}=r^{-2},\quad g^{zz}=1.
\end{align}
Non-vanishing components of the Levi-Civit\'a connection on ${\cal M}_+$ are
\begin{align}
\Gamma^T_{r\Phi}=-\frac{m}{r},\quad \Gamma^\Phi_{r\Phi}=\frac{1}{r},\quad \Gamma^r_{\Phi\Phi}=-\frac{r}{{\cal C}^2}. 
\end{align}
The exterior metric (\ref{Lewis}) describes a spinning cosmic string~\cite{Deser:1983tn,Jensen:1992wj} and admits a closed timelike curve in the region $r<|m|$ where $v^\mu(\partial/\partial x^\mu)=\partial/\partial\Phi$ is timelike.

The dynamics of a thin-shell $\Sigma$ as a matching hypersurface between ${\cal M}_-$ and ${\cal M}_+$ has been investigated for timelike $\Sigma$ in~\cite{Mena:2007dy} and for null $\Sigma$ in~\cite{Khakshournia:2011cj}.
Here we follow the argument in~\cite{Khakshournia:2011cj} in more detail.

In order to identify the description of the null shell $\Sigma$, we first study the most general affinely parametrized ingoing null geodesic $\gamma$ described by $x_+^\mu=x_+^\mu(\lambda)$ with an affine parameter $\lambda$ in the exterior spacetime ${\cal M}_+$, of which tangent vector is given by $k^\mu(=\D x_+^\mu/\D\lambda)$.
Since the exterior spacetime (\ref{Lewis}) admits the following Killing vectors
\begin{align}
\xi_1^\mu\frac{\partial}{\partial x^\mu}=\frac{\partial}{\partial T},\qquad \xi_2^\mu\frac{\partial}{\partial x^\mu}=\frac{\partial}{\partial \Phi},\qquad \xi_3^\mu\frac{\partial}{\partial x^\mu}=\frac{\partial}{\partial z},
\end{align}
there are three conserved quantities $E:=-k_\mu \xi_1^\mu$, $K:=k_\mu \xi_2^\mu$, and $V_z:=k_\mu \xi_3^\mu$ along $\gamma$, which give
\begin{align}
k^T=\frac{E(r^2-m^2) - mK}{r^2},\qquad k^\Phi=\frac{mE + K}{r^2},\qquad k^z=V_z.\label{T-Phi}
\end{align}
We assume $E>0$ so that $\gamma$ is future directed in a far region $r\to \infty$.
Then, the null condition $\D s_+^2=0$ along $\gamma$ with Eq.~(\ref{T-Phi}) gives the following master equation for $r(\lambda)$:
\begin{align}
\biggl(\frac{\D r}{\D\lambda}\biggl)^2=\frac{(E^2-V_z^2)(r^2-b^2)}{{\cal C}^2r^2}~~\to~~k^r=\frac{\D r}{\D\lambda}=-\frac{\sqrt{(E^2-V_z^2)(r^2-b^2)}}{{\cal C}r},\label{master-r}
\end{align}
where the minus sign is taken for ingoing $\gamma$ and $b$ is defined by
\begin{align}
b:=\frac{K+mE}{\sqrt{E^2-V_z^2}}\label{def-b}
\end{align}
for $E^2\ne V_z^2$.
The master equation (\ref{master-r}) requires $E^2-V_z^2\ge 0$ and shows that $r=|b|$ is the turning radius.
The general solution of the master equation for $E^2\ne V_z^2$ is given by 
\begin{align}
&\lambda-\lambda_0=-{\cal C}\sqrt{\frac{r^2-b^2}{E^2-V_z^2}}~~\to~~r(\lambda)^2=\frac{E^2-V_z^2}{{\cal C}^2}(\lambda-\lambda_0)^2+b^2,\label{r-lambda}
\end{align}
where $\lambda_0$ is an integration constant and $\lambda=\lambda_0$ corresponds to the turning radius $r=b$.
The null geodesic $\gamma$ enters the region with closed timelike curves for $|b|< |m|$, or equivalently $|K+mE|< |m|\sqrt{E^2-V_z^2}$.

Using Eqs.~(\ref{T-Phi}) and (\ref{master-r}), we obtain
\begin{align}
&\frac{\D T}{\D r}=-\frac{{\cal C}(Er^2-mb\sqrt{E^2-V_z^2})}{r\sqrt{(E^2-V_z^2)(r^2-b^2)}},\label{geo-lewis1}\\
&\frac{\D \Phi}{\D r}=-\frac{b{\cal C}}{r\sqrt{r^2-b^2}},\\
&\frac{\D z}{\D r}=-\frac{V_z{\cal C}r}{\sqrt{(E^2-V_z^2)(r^2-b^2)}}\label{geo-lewis3}
\end{align}
along $\gamma$.
We define three functions as
\begin{align}
&v(T,r):=T+r_*(r),\qquad \psi(\Phi,r):=\Phi+r_\Phi(r),\qquad Z(z,r):=z+r_z(r),
\end{align}
where 
\begin{align}
&r_*(r):=\int \frac{{\cal C}(Er^2-mb\sqrt{E^2-V_z^2})}{r\sqrt{(E^2-V_z^2)(r^2-b^2)}}\D r,\\
&r_\Phi(r):=\int\frac{b{\cal C}}{r\sqrt{r^2-b^2}}\D r,\\
&r_z(r):=\int\frac{V_z{\cal C}r}{\sqrt{(E^2-V_z^2)(r^2-b^2)}}\D r, 
\end{align}
and then $v$, $\psi$, and $Z$ are constant along $\gamma$ by Eqs.~(\ref{geo-lewis1})--(\ref{geo-lewis3}).
By coordinate transformations $T=v-r_*(r)$, $\Phi=\psi-r_\Phi(r)$, and $z=Z-r_z(r)$, the exterior metric (\ref{Lewis}) becomes
\begin{align}
\D s_+^2=&-\D v\biggl(\D v-\frac{2{\cal C}Er}{\sqrt{(E^2-V_z^2)(r^2-b^2)}}\D r+2m\D \psi\biggl) \nonumber\\
&-\frac{2{\cal C}Kr}{\sqrt{(E^2-V_z^2)(r^2-b^2)}}\D r\D \psi-\frac{2V_z{\cal C}r}{\sqrt{(E^2-V_z^2)(r^2-b^2)}}\D r\D Z \nonumber\\
&+(r^2-m^2)\D \psi^2+\D Z^2 \label{Lewis2}
\end{align}
in the coordinates $(v,r,\psi,Z)$.
Now we consider a hypersurface given by $v=v_0=$constant.
Its normal vector $l_\mu\D x^\mu=-(\nabla_\mu v)\D x^\mu=-\D v$ satisfies 
\begin{align}
l_\mu l^\mu=g^{vv}=-\frac{V_z^2(r^2-m^2) + K^2}{(K+mE)^2 - (E^2-V_z^2)r^2}
\end{align}
and therefore $v=v_0$ is a null hypersurface for $K=V_z=0$.

Hence, we identify $k^\mu$ with $K=0$ and $V_z=0$ (and then $b=m$ and $Z=z$) as the tangent vector to the generators of the null shell $\Sigma$ described by $v=v_0$.
Then, the exterior metric (\ref{Lewis}) and $k^\mu$ are given in the coordinates $(v,r,\psi,z)$ as
\begin{align}
&\D s_+^2=-\D v\biggl(\D v-\frac{2{\cal C}r}{\sqrt{r^2-m^2}}\D r+2m\D \psi\biggl)+(r^2-m^2)\D \psi^2+\D z^2, \label{Lewis2-0}\\
&k^\mu\frac{\partial}{\partial x^\mu}=-\frac{\sqrt{r^2-m^2}}{{\cal C}r}\frac{\partial}{\partial r},
\end{align}
where we have set $E=1$ by an affine transformation $E\lambda\to \lambda$.
Since $\psi$ and $z(=Z)$ are constant along the generators, we shall install intrinsic coordinates $y^a=(\lambda,\theta^A)$ on $\Sigma$ as $\theta^A\equiv (\psi,z)$, where $\lambda=\lambda(r)$ is given by Eq.~(\ref{r-lambda}) with $E=1$, $V_z=0$, and $b=m$.
Then, $e^\mu_a:=\D x^\mu/\D y^a$ are given by 
\begin{align}
&e^\mu_\lambda\frac{\partial}{\partial x^\mu}=k^\mu\frac{\partial}{\partial x^\mu},\qquad e^\mu_\psi\frac{\partial}{\partial x^\mu}=\frac{\partial}{\partial \psi},\qquad e^\mu_z\frac{\partial}{\partial x^\mu}=\frac{\partial}{\partial z},
\end{align}
which satisfy $k_\mu k^\mu=0$ and $k_\mu e^\mu_A=0$.
The induced metric on $\Sigma$ is given by 
\begin{align}
\D s_{\Sigma+}^2=\sigma_{AB}^+\D \theta^A\D\theta^B=&(r^2-m^2)|_\Sigma\D \psi^2+\D z^2={\cal C}^{-2}(\lambda-\lambda_0)^2\D \psi^2+\D z^2. \label{Lewis2}
\end{align}
The transverse null vector completing the basis is given by 
\begin{align}
N_\mu\D x^\mu=-\frac{r^2}{2(r^2-m^2)}\D v+\frac{{\cal C}r}{\sqrt{r^2-m^2}}\D r,
\end{align}
which satisfies $N_\mu N^\mu=0$, $N_\mu k^\mu=-1$, and $N_\mu e^\mu_A=0$.
Then, from Eq.~(\ref{Cab-def-original}), non-vanishing components of the transverse curvature of $\Sigma$ are computed to give
\begin{align}
\label{C+-Lewis}
\begin{aligned}
C^+_{\lambda\psi}=&C^+_{\psi\lambda}=\frac{m}{{\cal C}\sqrt{r^2-m^2}}\biggl|_\Sigma=-\frac{m}{\lambda-\lambda_0},\\
C^+_{\psi\psi}=&\frac{r^2}{2{\cal C}\sqrt{r^2-m^2}}\biggl|_\Sigma=-\frac{(\lambda-\lambda_0)^2+m^2{\cal C}^2}{2{\cal C}^2(\lambda-\lambda_0)}.
\end{aligned}
\end{align}

By setting $m=0$ and ${\cal C}=1$ and replacing $(T,r,\Phi)$ by $(t,\rho,\varphi)$, we can obtain the results on ${\cal M}_-$ described by the flat metric (\ref{flat}).
Equation~(\ref{r-lambda}) with $V_z=0$ and $b=0$ gives
\begin{align}
&\rho=-{\bar E}({\lambda}-{\lambda}_0),\label{r-lambda2}
\end{align}
where we have used the same constant $\lambda_0$ as that in ${\cal M}_+$ without loss of generality by an affine transformation $\lambda\to \lambda+b$.
${\bar E}$ is a constant related to the conserved quantity along a null generator of $\Sigma$ associated with the Killing vector $\xi^\mu(\partial/\partial x^\mu)=\partial/\partial t$.
As seen from ${\cal M}_-$, $\lambda=\lambda_0$ corresponds to the axis of symmetry $\rho=0$.
From Eqs.~(\ref{Lewis2}), we obtain the induced metric on $\Sigma$ as
\begin{align}
\D s_{\Sigma-}^2=\sigma^-_{AB}\D \theta^A\D\theta^B=&\rho^2|_\Sigma\D \psi^2+\D z^2={\bar E}^2(\lambda-\lambda_0)^2\D \psi^2+\D z^2. \label{Lewis2-}
\end{align}
From Eq.~(\ref{C+-Lewis}), we obtain the transverse curvature of $\Sigma$ as
\begin{align}
C^-_{\lambda\psi}=C^-_{\psi\lambda}=0,\qquad C^-_{\psi\psi}=\frac{\rho}{2{\bar E}}\biggl|_\Sigma=-\frac12(\lambda-\lambda_0).
\end{align}
By the first junction conditions $[\sigma_{AB}]=0$, we obtain
\begin{align}
{\bar E}={\cal C}^{-1}.
\end{align}
Then, using the Barrab\`{e}s-Israel junction conditions (\ref{components3-n2}) and 
\begin{align}
&[C_{\lambda\lambda}]=0,\qquad [C_{\lambda\psi}]=[C_{\psi\lambda}]=-\frac{m}{\lambda-\lambda_0},\\
&[C_{\psi\psi}]=-\frac{1}{2(\lambda-\lambda_0)}\biggl\{\frac{1-{\cal C}^2}{{\cal C}^2}(\lambda-\lambda_0)^2+m^2\biggl\},
\end{align}
we obtain
\begin{align}
\mu=&\frac{(1-{\cal C}^2)(\lambda-\lambda_0)^2+m^2{\cal C}^2}{2\kappa_4(\lambda-\lambda_0)^3},\quad j^\psi=-\frac{m{\cal C}^2}{\kappa_4(\lambda-\lambda_0)^3},\quad j^z=0,\quad p=0.
\end{align}
Using $\lambda-\lambda_0=-{\cal C}\sqrt{r^2-m^2}$ from Eq.~(\ref{r-lambda}), we can write those quantities in terms of $r$ as
\begin{align}
\mu=&\frac{({\cal C}^2-1)r^2-{\cal C}^2m^2}{2\kappa_4{\cal C}(r^2-m^2)^{3/2}}\biggl|_\Sigma,\quad j^\psi=\frac{m}{\kappa_4{\cal C}(r^2-m^2)^{3/2}}\biggl|_\Sigma,\quad j^z=0,\quad p=0.
\end{align}
This result is consistent with the one in~\cite{Khakshournia:2011cj}.
Since we have $p=0$ and $J\ne 0$, the induced energy-momentum tensor $t_{\mu\nu}$ on the null shell is of the Hawking-Ellis type III and violates all the standard energy conditions by Proposition~\ref{Prop:EC-null}.
For this reason, the present spacetime cannot be a model describing gravitational collapse of any kind of physically reasonable matter field.

Our result is conclusive because we have not used any approximation.
$t_{\mu\nu}$ on the shell is now given by 
\begin{align}
t^{\mu\nu}=&(-k_{\eta}u^\eta)^{-1}[\mu k^\mu k^\nu+j^\psi(k^\mu e^\nu_\psi+e^\mu_\psi k^\nu)]. \label{t-III2}
\end{align}
For our interest, let us see what happens in the slow-rotation approximation up to the linear order of $m/r$.
In this approximation, Eq.~(\ref{t-III2}) can be written as 
\begin{align}
&t^{\mu\nu}\simeq (-k_{\eta}u^\eta)^{-1}\mu \ell^\mu\ell^\nu, \label{t-II2}\\
&\ell^{\mu}:=k^\mu+\frac{j^\psi}{\mu} e^\mu_\psi.
\end{align}
The components of $\ell^{\mu}$ in the coordinates $(v,r,\psi,z)$ are
\begin{align}
\ell^{\mu}\frac{\partial}{\partial x^\mu}=&-\frac{\sqrt{r^2-m^2}}{{\cal C}r}\frac{\partial}{\partial r}+\frac{2m}{({\cal C}^2-1)r^2-{\cal C}^2m^2}\frac{\partial}{\partial\psi}
\end{align}
and its squared norm is given by
\begin{align}
&\ell_\mu\ell^{\mu}=\frac{4m^2(r^2-m^2)}{[({\cal C}^2-1)r^2-{\cal C}^2m^2]^2}\simeq 0.
\end{align}
Since $\ell^\mu$ is null in the slow-rotation approximation, one could misunderstand from Eq.~(\ref{t-II2}) that $t^{\mu\nu}$ is of the Hawking-Ellis type II and all the standard energy conditions are satisfied (violated) for $\mu\ge (<)0$.
Of course, it is a wrong conclusion caused by the approximation.

\subsection{Cosmological phase transition}

As the last application, this subsection considers a sudden transition of the universe from the anisotropic Bianchi I expansion to the isotropic flat Friedmann-Lema\^{\i}tre-Robertson-Walker (FLRW) expansion.
We consider the past spacetime ${\cal M}_-$ described by the following Bianchi-I metric
\begin{align}
&\D s_-^2=-\D {t}^2+a(t)^2(\D x^2+\D y^2)+\D z_-^2
\end{align}
in the coordinates $x_-^\mu=(t,x,y,z_-)$ and the future spacetime ${\cal M}_+$ described by the flat FLRW metric
\begin{align}
&\D s_+^2=-\D {t}^2+a(t)^2(\D x^2+\D y^2+\D z_+^2)
\end{align}
in the coordinates $x_+^\mu=(t,x,y,z_+)$ with the same scale factor $a(t)$.
We assume that the scale factor $a(t)$ is the same both in ${\cal M}_-$ and ${\cal M}_+$ as a solution to the Einstein equations $G_{\mu\nu}+\Lambda g_{\mu\nu}=\kappa_4 T_{\mu\nu}$. 
As the Einstein tensor in ${\cal M}_-$ and ${\cal M}_+$ are given by 
\begin{align}
&{G^\mu}_\nu|_-=\mbox{diag}\biggl(-\frac{{\dot a}^2}{a^2},-\frac{{\ddot a}}{a},-\frac{{\ddot a}}{a},-\frac{{\dot a}^2+2a{\ddot a}}{a^2}\biggl),\\
&{G^\mu}_\nu|_+=\mbox{diag}\biggl(-\frac{3{\dot a}^2}{a^2},-\frac{{\dot a}^2+2a{\ddot a}}{a^2},-\frac{{\dot a}^2+2a{\ddot a}}{a^2},-\frac{{\dot a}^2+2a{\ddot a}}{a^2}\biggl),
\end{align}
the matter fields in ${\cal M}_-$ and ${\cal M}_+$ are different.
In~\cite{Poisson:2002nv}, the author assumed $a(t)=(t/t_0)^{1/2}$ and $\Lambda=0$, where $t_0$ is a constant.
Then, we have 
\begin{align}
&{G^\mu}_\nu|_-=\mbox{diag}\biggl(-\frac{1}{4t^2},\frac{1}{4t^2},\frac{1}{4t^2},\frac{1}{4t^2}\biggl),\\
&{G^\mu}_\nu|_+=\mbox{diag}\biggl(-\frac{3}{4t^2},\frac{1}{4t^2},\frac{1}{4t^2},\frac{1}{4t^2}\biggl).
\end{align}
In this case, a matter field in ${\cal M}_-$ may be a stiff fluid, namely a perfect fluid obeying an equation of state $p=\rho$, while a matter field in ${\cal M}_+$ may be a radiation fluid, namely a perfect fluid obeying an equation of state $p=\rho/3$.
Hereafter we keep $a(t)$ arbitrary.

\subsubsection{Transition at spacelike $\Sigma$}

First, we consider the case where the transition occurs at a spacelike hypersurface $\Sigma$ given by $t=t_0=$constant both on ${\cal M}_-$ and ${\cal M}_+$ and then the first junction conditions require $a(t_0)^2=1$.
The induced metric on $\Sigma$ is given by 
\begin{align}
&\D s_{\Sigma}^2=h_{ab}\D y^a\D y^b=\D x^2+\D y^2+\D z^2,
\end{align}
where we have installed intrinsic coordinates $y^a=(x,y,z)$ on $\Sigma$ and the parametric equations $x^\mu=x^\mu(y^a)$ describing $\Sigma$ are given by
\begin{align}
&t=t_0,\quad x=x,\quad y=y,\quad z_-=z~~\mbox{on}~~{\cal M}_-,\\
&t=t_0,\quad x=x,\quad y=y,\quad z_+=z~~\mbox{on}~~{\cal M}_+.
\end{align}
The unit normal vector $n^\mu$ to $\Sigma$ is given by 
\begin{align}
n^\mu\frac{\partial}{\partial x^\mu}=\frac{\partial}{\partial t}
\end{align}
both on ${\cal M}_-$ and ${\cal M}_+$.
Note that the timelike vector $n^\mu$ points in an increasing direction of $t$, which is consistent with the assumption in Sec.~\ref{sec:non-null} that $n^\mu$ points from ${\cal M}_-$ to ${\cal M}_+$.

Using $e^\mu_a=\delta^\mu_a$ in Eq.~(\ref{K-def}), we obtain $K_{ab}= \Gamma^0_{ab}$ and therefore non-zero components of $K_{ab}$ on ${\cal M}_-$ and ${\cal M}_+$ are given by 
\begin{align}
&K^-_{xx}=K^-_{yy}=a{\dot a}|_{t=t_0},\\
&K^+_{xx}=K^+_{yy}=K^+_{zz}=a{\dot a}|_{t=t_0},
\end{align}
which give
\begin{align}
K^-=K^-_{ab}h^{ab}=\frac{2{\dot a}}{a}\biggl|_{t=t_0},\qquad K^+=K^+_{ab}h^{ab}=\frac{3{\dot a}}{a}\biggl|_{t=t_0}.
\end{align}
Thus, $[K_{ab}]$ admits only a single non-zero component $[K_{zz}]=a{\dot a}|_{t=t_0}$.
From the Israel junction conditions (\ref{j-eq-summary1}) with $\varepsilon=-1$ and $[K]={\dot a}/a|_{t=t_0}$, we obtain non-zero components of $t_{ab}$ as
\begin{align}
\kappa_4t_{xx}=\kappa_4t_{yy}=-a{\dot a}|_{t=t_0}.
\end{align}
Since eigenvalues of $t_{ab}v^{b}=\lambda h_{ab}v^{b}$ are given by $\lambda=\{0,-{\dot a}/(\kappa_4a)|_{t=t_0}\}$, the DEC is violated and the NEC, WEC, and SEC are satisfied (violated) for ${\dot a}\le (>)0$ by Proposition~\ref{Prop:EC-spacelike}.

\subsubsection{Transition at null $\Sigma$}

Next, we consider the case where the transition occurs at a null hypersurface $\Sigma$.
The system with $a(t)=(t/t_0)^{1/2}$ and $\Lambda=0$ has been studied in~\cite{Poisson:2002nv} as a simplified version of the example in~\cite{Barrabes:1998rp}, however, it contains an error that changes the conclusion.

As seen from ${\cal M}_-$, we describe the transition null hypersurface $\Sigma$ by $t-z_-=$constant.
The null vector $k^\mu=-g^{\mu\nu}\nabla_\nu (t-z_-)$ is computed to give
\begin{align}
k^\mu\frac{\partial}{\partial x^\mu}=\frac{\partial}{\partial t}+\frac{\partial}{\partial z_-},
\end{align}
which satisfies $k^\nu\nabla_\nu k^\mu=0$.
Hence, $k^\mu$ is tangent to the null generators of $\Sigma$ which are parametrized by an affine parameter $\lambda$.
By $k^t=1$, $t$ is an affine parameter on this side of $\Sigma$.
Since $x$ and $y$ are constant on the generators, we install intrinsic coordinates $y^a=(\lambda,\theta^A)$ on $\Sigma$ as $\lambda=t$ and $\theta^A=(x,y)$, and then $e^\mu_a:=\partial x^\mu/\partial y^a$ are given by 
\begin{align}
e^\mu_\lambda=k^\mu,\qquad e^\mu_A=\delta^\mu_A,
\end{align}
which satisfies $k_\mu e^\mu_A=0$ and $g_{\mu\nu}e^\mu_A e^\mu_B=a^2\delta_{AB}$.
From Eq.~(\ref{line-null}), we obtain the induced metric on $\Sigma$ as 
\begin{align}
\D s_{\Sigma-}^2=\sigma_{AB}^-\D\theta^A\D\theta^B=a(\lambda)^2(\D x^2+\D y^2).
\end{align}
The transverse null vector completing the basis is given by 
\begin{align}
N_\mu\D x^\mu=-\frac12(\D t+\D z_-),
\end{align}
which satisfies $N_\mu N^\mu=0$, $N_\mu k^\mu=-1$, and $N_\mu e^\mu_A=0$.
From Eq.~(\ref{Cab-def-original}), non-vanishing components of the transverse curvature of $\Sigma$ are computed to give
\begin{align}
C^-_{AB}=\frac{\dot a}{2a}\sigma_{AB}^-\biggl|_{t=\lambda},
\end{align}
where a dot denotes differentiation with respect to $t$.

As seen from ${\cal M}_+$, we describe $\Sigma$ by $\int a^{-1}\D t-z_+=$constant, which is obtained by integrating $\D t=a(t)\D z_+$.
The null vector $k^\mu=-g^{\mu\nu}\nabla_\nu (\int a^{-1}\D t-z_+)$ tangent to the null generators of $\Sigma$ is given by\footnote{In~\cite{Poisson:2002nv}, $k^\mu$ was erroneously identified as $k^\mu\partial/\partial x^\mu=\partial/\partial t+a^{-1}\partial/\partial z_+$, which affected the following argument.} 
\begin{align}
k^\mu\frac{\partial}{\partial x^\mu}=a^{-1}\frac{\partial}{\partial t}+a^{-2}\frac{\partial}{\partial z_+},
\end{align}
which satisfies $k^\nu\nabla_\nu k^\mu=0$.
Hence, the null generators of $\Sigma$ are parametrized by an affine parameter $\lambda$ also on this side of $\Sigma$.
However, $t$ is {\it not} an affine parameter on this side and $k^t=\D t/\D \lambda=a(t)^{-1}$ is integrated to give $\lambda=\int a\D t$.
We write the inverse function of $\lambda=\int a\D t$ as $t=t_+(\lambda)$.
Since $x$ and $y$ are constant on the generators, we install intrinsic coordinates $y^a=(\lambda,\theta^A)$ on $\Sigma$ as $\lambda=\int a\D t$ and $\theta^A=(x,y)$, and then $e^\mu_a:=\partial x^\mu/\partial y^a$ are given by 
\begin{align}
e^\mu_\lambda=k^\mu,\qquad e^\mu_A=\delta^\mu_A,
\end{align}
which satisfies $k_\mu e^\mu_A=0$ and $g_{\mu\nu}e^\mu_A e^\mu_B=a^2\delta_{AB}$.
From Eq.~(\ref{line-null}), we obtain the induced metric on $\Sigma$ as 
\begin{align}
\D s_{\Sigma+}^2=\sigma_{AB}^+\D\theta^A\D\theta^B=a(t)^2|_{t=t_+(\lambda)}(\D x^2+\D y^2).
\end{align}
Since the coordinate $t$ is the same on ${\cal M}_-$ and ${\cal M}_+$, the first junction conditions $[\sigma_{AB}]=0$ are satisfied.
The transverse vector completing the basis is 
\begin{align}
N_\mu\D x^\mu=-\frac12(a\D t+a^2\D z_+),
\end{align}
which satisfies $N_\mu N^\mu=0$, $N_\mu k^\mu=-1$, and $N_\mu e^\mu_A=0$.
From Eq.~(\ref{Cab-def-original}), non-vanishing components of the transverse curvature of $\Sigma$ are computed to give
\begin{align}
C^+_{AB}=\frac{1}{2}{\dot a}\sigma_{AB}^+\biggl|_{t=t_+(\lambda)}.
\end{align}

The jump of the transverse curvature at $\Sigma$ is obtained in terms of $t$ as
\begin{align}
[C_{\lambda\lambda}]=[C_{\lambda B}]=0,\qquad [C_{AB}]=\frac12\biggl({\dot a}|_{t=t_+(\lambda)}-\frac{{\dot a}}{a}\biggl|_{t=\lambda}\biggl)\sigma_{AB}=\frac{{\dot a}}{2}\biggl(1-\frac{1}{a}\biggl)\sigma_{AB},
\end{align}
where $\sigma_{AB}\equiv \sigma_{AB}^+(=\sigma_{AB}^-)$.
Then, the Barrab\`{e}s-Israel junction conditions (\ref{components3-n2}) give
\begin{align}
\kappa_4\mu=&-{\dot a}\biggl(1-\frac{1}{a}\biggl),\qquad j^A=0,\qquad p=0.
\end{align}
As we have $\mu\ne 0$, $p=J=0$, the induced energy-momentum tensor $t_{\mu\nu}$ on the null shell is of the Hawking-Ellis type II and all the standard energy conditions are satisfied (violated) for ${\dot a}(1-a^{-1})\le (>)0$ by Proposition~\ref{Prop:EC-null}.
Thus, in the expanding universe with ${\dot a}>0$, such a phase transition of the cosmic expansion at null $\Sigma$ is possible without violating any of the standard energy conditions if it occurs in the early stages of the universe where $a\le 1$ holds.


\section*{Acknowledgments}
The author is very grateful to Max-Planck-Institut f\"ur Gravitationsphysik (Albert-Einstein-Institut) for hospitality and support, where a large part of this work was carried out.


\end{document}